\documentclass{article}

\usepackage{arxiv}

\usepackage[utf8]{inputenc} 
\usepackage[T1]{fontenc}    
\usepackage{hyperref}       
\usepackage{url}            
\usepackage{booktabs}       
\usepackage{amsfonts}       
\usepackage{nicefrac}       
\usepackage{microtype}      
\usepackage{lipsum}		
\usepackage{graphicx}
\usepackage{natbib}
\usepackage{doi}
\usepackage{amsmath}
\usepackage{tikz}
\usepackage{algorithm}
\title{Optimal Experimental Design for Network Experiments under Interference}

\author{{\hspace{1mm}Zuhra F. S. Lebbe}\\
	Department of Mathematics and Statistics\\
	Texas Tech University\\
	Lubbock, 79409, Texas, United States \\
	\texttt{zseleima@ttu.edu} \\
	\And
	{\hspace{1mm}Asim K. Dey}\\
	Department of Mathematics and Statistics\\
	Texas Tech University\\
	Lubbock, 79409, Texas, United States \\
	\texttt{a.dey@ttu.edu} \\
}

\hypersetup{
  hidelinks,
  pdftitle={Optimal Experimental Design for Network Experiments under Interference},
  pdfauthor={Zuhra F. S. Lebbe, Asim K. Dey},
  pdfsubject={Experimental design under network interference},
  pdfkeywords={complex networks, experimental design, interference, heterogeneous causal treatment effects}
}

\begin{document}
\maketitle

\begin{abstract}
	Experimental design under network interference is challenging because outcomes may depend on the treatment assignments of neighboring units. Existing approaches account for network structure but are typically assessed on small or simplified networks, limiting their applicability to complex real-world settings. We propose a network-aware treatment allocation framework that jointly accounts for allocation balance and network topology via an optimality criterion based on the Fisher information matrix. To address the resulting combinatorial optimization problem, we develop an efficient local search algorithm that scales to large networks. We further study the causal properties of the resulting designs by examining the estimation of total, direct, and indirect treatment effects in the presence of interference. Simulation studies across a range of random graph models, including Erd\H{o}s--R\'enyi, geometric random graphs, preferential attachment, and stochastic block models, illustrate how network topology influences optimal treatment allocations. Applications to college housing and ego-Facebook networks demonstrate the practical advantages of topology-aware experimental designs.
\end{abstract}


\section{Introduction}
Modern research in epidemiology, social media, communication, and online platforms increasingly focuses on how treatment assignment strategies should be designed and implemented within populations. 
Here, treatment assignment refers to the allocation of experimental interventions, such as resources, policies, or medical treatments, across population units. A large literature on optimal treatment allocation and experimental design has been developed under the classical assumption of the \textit{Stable Unit Treatment Value Assumption} (SUTVA)~\citep{Rubin_1974,cox1958planning}, under which a unit's potential outcome is assumed to be unaffected by others' treatments. Within this framework, optimality is typically defined in terms of variance reduction, efficiency gains, or improved covariate balance for estimators of average treatment effects~\citep{MorganRubin2012, Johansson2021,Ravichandran2023, Azriel2021,li2023optimal,zhang2025adaptivedesignoptimizingtreatment}. 

In many real-world applications, the optimal design of treatment allocation is challenging due to the complex dependence induced by social and spatial structures, as well as cross-unit effects, whereby a unit's outcome may depend not only on its own treatment but also on the treatments received by neighboring units. Such cross-unit interactions, often referred to as \textit{interference}, arise naturally through social, spatial, and physical mechanisms and undermine classical SUTVA assumptions by violating the premise that potential outcomes for population units are statistically independent under treatment assignments. A motivating example comes from the early stages of the COVID‑19 pandemic, when health authorities had to allocate scarce vaccines, testing kits, and therapeutics across large, interconnected populations. Because transmission occurred through social contact networks, treating or vaccinating particular individuals reduced risk for others through indirect protection along their contacts (network ties). As a result, network structure became a central consideration for allocation decisions.

These considerations motivate a growing line of research on designing experiments, particularly on \textit{networks}, where dependence among population units is explicitly structured by an underlying graph encoding social, spatial, or contact relationships. In networked settings, treatment assignments propagate effects along edges, and both direct and indirect effects depend on the topology of the network and the placement of treated units within it. Consequently, experimental design needs to account not only for the number of treated units but also for their positions in the network, raising new questions about optimal allocation. Designing experiments on networks, therefore, requires methodologies that jointly consider treatment assignment, network structure, and interference patterns, departing fundamentally from classical designs developed under independence assumptions.

A growing methodological literature has addressed these challenges by developing formal frameworks for the design of experiments on networks. \cite{Wen-Hsiu_2023} propose an optimal design framework for undirected networks that accounts for both direct treatment effects and indirect effects. \cite{Parker2016} develop an experimental design method specifically for connected nodes in social networks. 
They extended this work by proposing an optimal block design by defining blocks that account for the interrelations of groups of experimental units within a network~\citep{Koutra_2021}. 
\cite{Zhang2025} proposes rerandomization algorithm that generate experimental designs satisfying balance conditions derived from optimality criteria under specific network assumptions. \cite{ugander2013} introduce graph cluster randomization, providing an efficient algorithm to compute exposure probabilities for units under various network exposure conditions. 
\cite{Pokhilko2019} introduce an optimal design approach for A/B testing in network settings using a conditional auto-regressive (CAR) model, optimizing treatment assignments to improve estimation precision under interference. \cite{Basse2018} propose a model-assisted experimental design framework that uses pre-treatment network information to construct restricted randomization schemes, improving estimation efficiency by accounting for network-induced correlation in outcomes. \cite{TaylorEckles2018} survey the broader design space, defining units, interventions, randomization mappings (over nodes or edges), and Fisherian randomization inference, and emphasize how analysis must account for network dependence. 
\cite{Christina_2022} introduce graph-agnostic randomized designs under heterogeneous linear neighborhood interference, providing simple estimators for total, direct, and interference effects that leverage baseline measurements and remain unbiased without requiring the knowledge of the network structure. 
\cite{Gui2015} investigate network effects in large-scale online experiments and propose estimation methods for the average treatment effect that account for interference.
\cite{Li2022} focus on treatment effect estimation under network interference and develop consistent estimators in settings where standard approaches fail.
\cite{hu2022} provide a general, nonparametric framework for defining average direct and indirect effects under interference.

Although existing approaches offer valuable methodological contributions, they are predominantly evaluated on small-scale and simplified network structures. Consequently, their applicability to networks with the complexity, sparsity, and scale commonly observed in practice is not yet fully established. Motivated by these challenges, in this paper, we propose a network-aware design framework for treatment allocation in networked experiments. The framework builds on the optimality criterion of~\cite{Wen-Hsiu_2023}, which explicitly balances treatment allocation with network structure, and integrates it with a local search strategy to navigate the combinatorial design space. This combination allows the proposed approach to efficiently identify treatment assignments that account for network topology while maintaining computational feasibility for networks of practical size.

We further examine the causal treatment effects associated with the optimized treatment assignments, including the total treatment effect, average treatment effect, and average interference effect. The utility of the proposed approach is evaluated across a range of random graph models, including Erd\H{o}s--R\'enyi, geometric random graphs, preferential attachment models, and stochastic block models, as well as two real-world networks, namely a Facebook network and a college housing network.

The remainder of the article is organized as follows. Section \label{Methodology}~\ref{Sec:Methodology} introduces the proposed methodology, including the interference model and the optimal design framework for treatment assignment on networks. Section~\ref{Sec:Treatment effects} examines the causal treatment effects induced by the optimized treatment assignments. Section~\ref{Sec:Sim studies}  compares the proposed method with existing approaches and presents a comprehensive simulation study illustrating the proposed optimal design across a range of random network models under treatment-control and three-treatment settings. Section~\ref{Sec:real} applies the proposed design to two real-world networks. Section~\ref{Sec:discussion} concludes the paper with a discussion.

\section{Methodology}
\label{Sec:Methodology}

\subsection{Preliminaries and Problem Setup}
\label{Sec:Model1}
We consider an undirected network $G = (V, E)$, where $V$ is the set of nodes representing experimental units and $E$ is the set of edges encoding relationships among the nodes. Let $n = |V|$ denote the number of nodes, indexed by $i = 1, \ldots, n$. Suppose there are $m$ treatments $\{t_{(1)}, \ldots, t_{(m)}\}$. Let $t(i) \in \{t_{(1)}, \ldots, t_{(m)}\}$ denote the treatment assigned to node $i$. We assume that each node receives exactly one treatment.

We define the node-treatment incidence matrix as an $n \times m$ matrix $\boldsymbol{X} = (X_1, \ldots, X_n)^\top$, where each entry $\boldsymbol{X}_{ij}$ is defined as

\begin{equation*}\label{Eq:treatm}
\boldsymbol{X}_{ij}=\begin{cases} 1, & \mbox{if the
\(i^{th}\) node receives the \(j^{th}\) treatment},\\
0, & \mbox{otherwise},
\end{cases}
\end{equation*}

where $j = 1, \ldots, m$. 
That is, each row of $\boldsymbol{X}$ denoted by \( \boldsymbol{X}^T_i\), corresponds to node $i$ and contains exactly one non-zero entry. 
Given the treatment assignment matrix $\boldsymbol{X}$, we can model the observed responses across all nodes by the vector \(\boldsymbol{Y} = (y_1, \dots, y_n)^T \in \mathbb{R}^n\), where \(y_i\) represents the observed outcome for the $i^{th}$ node. We can model the relationship between treatments and responses as
\begin{equation}\label{Eq:linear}
    \boldsymbol{Y = X\beta + \epsilon},
\end{equation}

where $\boldsymbol{\beta} = (\beta_1,\dots, \beta_m)^T$ is the $m$-dimensional vector of treatment effects. The error term follows a multivariate normal distribution with mean zero and covariance matrix  $\sigma^2\boldsymbol{I_n}$, i.e., $\boldsymbol{\epsilon}\sim \mathcal{N}({\boldsymbol{0}, \sigma^2\boldsymbol{I_n}})$. The treatment effects $\boldsymbol{\beta}$ are treated as fixed but unknown parameters. 

Many network experiments depend on the core independence assumption SUTVA, also known as no interference~\citep{Rubin_1974,cox1958planning}. Under this assumption, the outcome for node $i$ depends only on its own treatment and is unaffected by the treatments assigned to its neighbors. In many epidemiological and social network settings, this independence fails because nodes influence one another through their connections. As a result, a node's response may correlate with its neighbors' responses and may also be shaped by the treatments its neighbors receive. This dependence is known as the \textit{network interference} effect~\citep{Ogburn_2017,EcklesKarrerUgander_2017,Ogburn02012024,Aronow_2017,viviano2025}.

In this study, we account for network interference effects arising from interactions among nodes. We define the $n \times n$ adjacency matrix $\boldsymbol{A}$ of the network $G$ as
\begin{equation*}\label{Eq:AD}
\boldsymbol{A}_{ik}=\begin{cases} 1, & \mbox{if nodes $i$ and $k$ are connected},\\
0, & \mbox{otherwise}.
\end{cases}
\end{equation*}

To exclude self-loops, we set \(\boldsymbol{A}_{ii}\) = 0. We assume that the network $G$ is undirected; so the adjacency matrix $\mathbf{A}$ is symmetric, i.e., \(\boldsymbol{A}_{ik} =\boldsymbol{A}_{ki}\) for all $i,k=1, 2,\ldots, n$.

Let $\boldsymbol{L}$ denote the \textit{treatment-treatment edge incidence matrix}, which summarizes how often pairs of treatments appear at the two ends of an edge in the network. Specifically, for $m$ treatments, we can define $\boldsymbol{L}$ by a
$m\times m$ matrix, whose $(j,\ell)$ entry is given by
\[
\boldsymbol{L}_{t(i)t(k)}=L_{j\ell}
=
\sum_{(i,k)\in E}
\mathbf{1}\{t(i)=t_{(j)},\ t(k)=t_{(\ell)}\},
\]

where $\mathbf{1}(\cdot)$ denotes the indicator function, $j,\ell=1,\ldots, m$. That is, $(j,\ell)$ entry counts the number of edges whose endpoints are assigned treatments $t_{(j)}$ and $t_{(\ell)}$, respectively. Since the network is undirected, $\boldsymbol{L}$ is symmetric, that is, \(\boldsymbol{L}_{t(i)t(k)}\) = \(\boldsymbol{L}_{t(k)t(i)}\). The treatment-treatment edge incidence matrix measures how treatments are distributed relative to the network structure and serves as a metric for the network interference effect.

\subsection{Network Model under Interference}
\label{Sec:Network Model}
We model network effects using a framework based on the \textit{random graph logistic model}, where the probability of an edge between two nodes depends on their treatment assignments. In this approach, the presence of an edge is modeled as a Bernoulli random variable, and its log odds are expressed as a linear function of node level covariates~\citep{wasserman1996logit,robins2007introduction,Clark_2022,Dey_2025A}. Specifically, the probability of an edge between nodes $i$ and $k$ can be written as
\begin{equation}\label{Eq:prob}
P(\boldsymbol{A}_{ik} = 1 \mid X_i, X_k) = \phi_{ik} = 
\frac{\exp(s_{ik} + \theta \boldsymbol{L}_{t(i)t(k)})}
{1 + \exp(s_{ik} + \theta \boldsymbol{L}_{t(i)t(k)})} ,
\end{equation}


where \(s_{ik} = \boldsymbol{X}^T_i \boldsymbol{X}_k\) measures the similarity between $i^{th}$ and $k^{th}$ nodes based on their covariates, and $\theta$ controls the strength and direction of how this similarity influences the probability of an edge. 
In this setting, the edges are assumed to be independent, i.e., $\boldsymbol{A}_{ik}$, are independent random variables. This probabilistic model for edge formation captures both local treatment similarity through $s_{ik}$, and global structural dependencies through $\boldsymbol{L}_{t(i)t(k)}$~\citep{zhang2020}.

Let $\boldsymbol{\alpha} = (\boldsymbol{\beta}^T, \sigma, \theta)^T$ denotes the vector of unknown parameters. Under the normality assumption of errors $\boldsymbol{\epsilon}$, i.e., $\boldsymbol{\epsilon}\sim \mathcal{N}({\boldsymbol{0}, \sigma^2\boldsymbol{I_n}})$, the conditional distribution of each response follows a normal distribution, i.e., 
$y_i \mid \boldsymbol{X}_i \sim \mathcal{N}(\boldsymbol{X}_i^T\boldsymbol{\beta}, \sigma^2)$.
As the network structure is observed, the conditional distribution of network edges follows a Bernoulli distribution with parameter $\phi_{ik}$, i.e., $\boldsymbol{A}_{ik} \mid (\boldsymbol{X}_i, \boldsymbol{X}_k) \sim Bernoulli(\phi_{ik})$. We assume that the marginal distribution of $\boldsymbol{A}_{ik}$ is independent of $\boldsymbol{\alpha}$. Under this setting, we can write the joint likelihood of $\boldsymbol{\alpha}$ as 
\begin{align*}
 \textit{L}(\boldsymbol{\alpha}) 
  &\propto P(\boldsymbol{Y, A} \mid \boldsymbol{X}) \nonumber \\
  &= P(\boldsymbol{Y} \mid \boldsymbol{X})P(\boldsymbol{A} \mid \boldsymbol{X}) \nonumber \\
  &=  \prod_{i=1}^n P(y_i \mid \boldsymbol{X}_i) \prod_{i\neq k}^n P( \boldsymbol{A}_{ik} \mid \boldsymbol{X}_i, \boldsymbol{X}_k) \nonumber \\
  &= \prod_{i=1}^n \frac{1}{\sqrt{2\pi\sigma^2}}\exp\left[-\frac{1}{2\sigma^2}(y_i-X_i^T\boldsymbol{\beta})^2\right]\prod_{i \neq k}^n \phi_{ik}^{\boldsymbol{A}_{ik}}(1-\phi_{ik})^{(1-\boldsymbol{A}_{ik})}.
\end{align*}

Based on the likelihood function, we can write the Fisher information matrix~\citep{Fisher1925, CasellaBerger2002}, which provides the information about $\boldsymbol{\alpha}$ that is contained in the observed data, as

\begin{align*}
\boldsymbol{J}(\boldsymbol{\alpha})
  &= - E[\nabla^2log \textit{L}(\boldsymbol{\alpha)}] \nonumber \\
  &= -E\begin{bmatrix}\begin{array}{cccccc}
 \frac{n_1}{\sigma^2} & 0 & \dots & 0 & 0 & 0 \\[4pt]
 0 & \frac{n_2}{\sigma^2} & \dots & 0 & 0 & 0 \\[4pt]
 \vdots & \vdots & \ddots & \vdots & \vdots & \vdots \\[4pt]
 0 & 0 & \dots & \frac{n_m}{\sigma^2} & 0 & 0 \\[4pt]
 0 & 0 & 0 & 0 & \frac{2n}{\sigma^2} & 0 \\[4pt]
 0 & 0 & 0 & 0 & 0 & \displaystyle\sum_{i=1}^n \sum_{k \neq i} 
 \boldsymbol{L}^2_{t(i)t(k)} \phi_{ik} (1 - \phi_{ik})
 \end{array}
 \end{bmatrix}
\end{align*}
where $n_j,~j=1, \ldots,m$, denotes the total number of nodes receiving the $j^{th}$ treatment. Note that $\sum_{j=1}^{m} n_j=n$, since each node receives exactly one of the $m$ possible treatments.

The Fisher information matrix provides a basis for establishing an optimal criterion. We define the optimal design as the one that maximizes the determinant of the Fisher information matrix, $\det(\boldsymbol{J}(\boldsymbol{\alpha}))$. Following \cite{Wen-Hsiu_2023}, we can write $\det(\boldsymbol{J}(\boldsymbol{\alpha}))$ as 
\begin{equation}\label{Eq:OptiCr}
\resizebox{0.9\linewidth}{!}{$
\begin{aligned}
\det(\boldsymbol{J}(\boldsymbol{\alpha}))=\tau = \prod_{j=1}^{m} n_j \left[\sum_{h=1}^m \binom{n_h}{2} \boldsymbol{L}^2_{hh}\frac{\exp(1 + \theta \boldsymbol{L}_{hh})}
{[1 + \exp(1 + \theta \boldsymbol{L}_{hh})]^2} + \sum_{j=1}^{m-1}\sum_{\ell>j}^m n_jn_{\ell}\boldsymbol{L}^2_{j\ell}\frac{\exp(\theta \boldsymbol{L}_{j\ell})}
{[1 + \exp(\theta \boldsymbol{L}_{j\ell})]^2} \right],
\end{aligned}
$}
\end{equation}

where the first component $\prod_{j=1}^{m} n_j$ promotes balance in treatment assignments, and the second component in the bracket captures the interaction between treatment groups and the underlying network structure. 
Here, $h = \min\{t(i),t(j)\}$, where $h,j,\ell=1,\ldots,m$. 

A constraint on $n_{j}$ is that $n_{j} > 0$ for all treatments $j$; in other words, every treatment must be assigned to at least one node. Otherwise, the determinant of the Fisher information matrix $det(\boldsymbol{J}(\boldsymbol{\alpha}))$ becomes zero. Based on the observed network structure, we can estimate the parameter $\theta$ using maximum likelihood estimation. However, in practice the second component in Eq.~\ref{Eq:OptiCr} is analytically intractable because it depends on the estimation of $\theta$. To simplify the analysis, we consider the limiting case in which $\theta$ approaches zero~\citep{Wen-Hsiu_2023}. The objective $\tau$ in Eq.~\ref{Eq:OptiCr} thus promotes efficient designs by utilizing both treatment balance and structural information embedded in the network's edge probabilities.


\subsection{Optimal Design on Networks under Interference}
\label{Sec:Optimal_Design}
To obtain the optimal design, we adapt an iterative local-exchange algorithm, originally introduced by \cite{Cook01081980} and refined for small-network settings by \cite{Koutra_2021}. We update the algorithm by combining the optimal criterion defined in Eq.~\ref{Eq:OptiCr}. The updated algorithm efficiently explores the space of possible treatment assignments in the network and finds the treatment allocation that maximizes the objective $\tau$. This is an approximate, yet heuristic and scalable, method for large network design optimization. The procedure proceeds as follows:

\begin{enumerate}
\item \textit{Step 1}: Initialize by generating a network $G = (V, E)$ where $V$ denotes the set of nodes such that $n = |V|$ and $E$ denotes the set of edges.

\item \textit{Step 2}: Consider two treatments $\{t_{(0)},t_{(1)}\}$. Generate an \textit{initial} random treatment assignment vector $\boldsymbol{t}=(t^1_{(0)},t^2_{(1)}, \ldots, ,t^n_{(0)})$ for the graph $G$, where each node is assigned exactly one treatment from $\{t_{(0)},t_{(1)}\}$. 

\item \textit{Step 3}: Compute the optimality criterion value $\tau$ (Using Eq.~\ref{Eq:OptiCr}) for this initial design.
\item \textit{Step 4}: Perform a local search to improve the design:
\begin{itemize}
\item For node $i$, $i= 1, \ldots, n$:
\begin{itemize}
\item Exchange the treatment assigned to node $i$ (say, $t_{(0)})$ with a different treatment (say, $t_{(1)}$). Obtain new treatment assignment vector $\boldsymbol{t}'$.

\item Recalculate the criterion, say, $\tau'$ (Using Eq.~\ref{Eq:OptiCr}) for the modified design $\boldsymbol{t}'$.
\item If the new value of $\tau'$ is higher than before, retain the change for $i$; otherwise, revert to $i$'s previous treatment assignment.
\end{itemize}
\end{itemize}

\item \textit{Step 5}: Repeat \textit{Step 4} until a complete pass through all $n$ nodes. 

\item \textit{Step 6}: Repeat \textit{Steps 2-5} for $R$ initial random treatment assignments. 

\item \textit{Step 7}: Return the highest observed criterion value for $R$ repetition denoted as $\tau^*$, with the corresponding treatment assignment $\boldsymbol{t}^*$.
\end{enumerate}

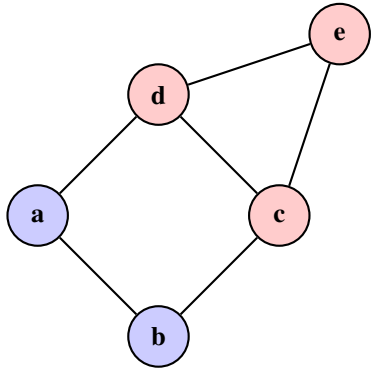
\begin{figure}[ht!]
\centering
\begin{tikzpicture}[scale=0.8,minimum size=0.8cm, thick]
 \tikzset{
    treat0/.style={circle, draw, fill=red!20,font=\bfseries},
    treat1/.style={circle, draw, fill=blue!20,font=\bfseries}
  }
    \node[treat1] (a) at (0,0) {a};
    \node[treat1] (b) at (2,-2) {b};
    \node[treat0] (c) at (4,0) {c};
    \node[treat0] (d) at (2,2) {d};
    \node[treat0] (e) at (5,3) {e};
    
    \draw (a) -- (b);
    \draw (a) -- (d);
    \draw (b) -- (c);
    \draw (c) -- (d);
    \draw (c) -- (e);
    \draw (d) -- (e);
\end{tikzpicture}
\caption{Network  $G^\text{toy}$ with two treatments.}
\label{fig:example_network}
\end{figure}

To illustrate the proposed framework, we present a toy example with a small undirected network $G^\text{toy}$ shown in Figure~\ref{fig:example_network}, consisting of 5 nodes and 6 edges. Two treatments, denoted as \( t_{(0)} \) and \( t_{(1)} \), are randomly assigned to the nodes and represented by color, red for \( t_{(0)} \) and blue for \( t_{(1)} \). That is, nodes \textit{a} and \textit{b} receive treatment \( t_{(1)} \), while nodes \textit{c}, \textit{d}, and \textit{e} receive treatment \( t_{(0)} \).


\newpage
For this case, we can write the treatment assignment matrix $\boldsymbol{X}$ as 
\[
\boldsymbol{X} =
\begin{array}{c@{\hspace{-0.3em}}c}
&
\begin{array}{cc}
t_{(0)} & t_{(1)}
\end{array}
\\
\begin{array}{c}
a \\ b \\ c \\ d \\ e
\end{array}
&
\left[
\begin{array}{cc}
0 & 1 \\
0 & 1 \\
1 & 0 \\
1 & 0 \\
1 & 0
\end{array}
\right]
\end{array}_{5 \times 2},
\]

where each row represents one node, and each column corresponds to a treatment. The entry $\boldsymbol{X}_{ij}=1$ if node $i$ receives treatment $j$ and 0 otherwise. We can write the adjacency matrix $\boldsymbol{A}$ of the network $G^\text{toy}$ as

\[
\boldsymbol{A} =
\begin{bmatrix}
0 & 1 & 0 & 1 & 0 \\
1 & 0 & 1 & 0 & 0 \\
0 & 1 & 0 & 1 & 1 \\
1 & 0 & 1 & 0 & 1 \\
0 & 0 & 1 & 1 & 0 
\end{bmatrix}_{5 \times 5},
\]

where $\boldsymbol{A}_{ik}=1$ indicates that nodes $i$ and $k$ are connected by an edge, while $\boldsymbol{A}_{ik}=0$ indicates no connection. Since the network is undirected, $\boldsymbol{A}$ is symmetric, and diagonal entries are zero as no node is connected to itself. 
The corresponding treatment-treatment edge incidence matrix $\boldsymbol{L}$ can be written as

\[
\boldsymbol{L} =
\begin{array}{c@{\hspace{-0.3em}}c}
&
\begin{array}{cc}
t_{(0)} & t_{(1)}
\end{array}
\\
\begin{array}{c}
t_{(0)} \\ t_{(1)}
\end{array}
&
\left[
\begin{array}{cc}
3 & 2 \\
2 & 1 \\
\end{array}
\right]
\end{array}_{2 \times 2}.
\]


That is, if node $i$ receives treatment $t_{(0)}$, so that $t(i)=t_{(0)}$, and node $k$ also receives treatment $t_{(0)}$, so that $t(k)=t_{(0)}$, then the 
treatment-treatment edge incidence matrix $\boldsymbol{L}$ yields 
$\boldsymbol{L}_{t(i)t(k)} = L_{t_{(0)}t_{(0)}} = 3$. This indicates that there are three edges whose two endpoints are both assigned treatment $t_{(0)}$. Similarly, two edges connecting nodes assigned treatments $t_{(0)}$ and $t_{(1)}$, and one edge between two nodes both assigned treatment $t_{(1)}$. The symmetry of the matrix $\boldsymbol{L}$ reflects the undirected nature of the graph.

To demonstrate how the optimization algorithm operates, we consider the network $G^\text{toy}$ with the initial random treatment assignment 
\resizebox{0.26\linewidth}{!}{$\boldsymbol{t}=(t^a_{(1)},t^b_{(1)},t^c_{(0)},t^d_{(0)},t^e_{(0)})$}. 
We first compute the optimality criterion value, denoted by $\tau$, using equation ~\ref{Eq:OptiCr}. 
Next, we apply the local search procedure: for a given node, we switch its treatment and evaluate whether this change improves $\tau$. 
For instance, let's consider node $a$, which receives treatment $t_{(1)}$.

\begin{itemize}
    \item We change its treatment from $t_{(1)}$ to $t_{(0)}$, resulting in a new treatment assignment $\boldsymbol{t'}=(t^a_{(0)},t^b_{(1)},t^c_{(0)},t^d_{(0)},t^e_{(0)})$ as presented in figure~\ref{fig:change of t}.
    
    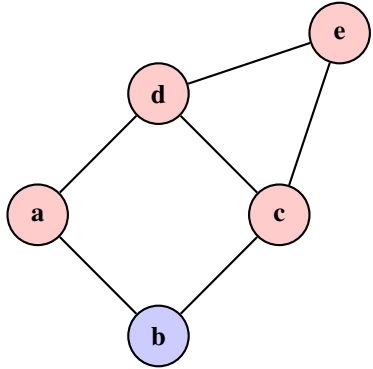
\begin{figure}[ht!]
\centering
\begin{tikzpicture}[scale=0.8,minimum size=0.8cm, thick]
 \tikzset{
    treat0/.style={circle, draw, fill=red!20,font=\bfseries},
    treat1/.style={circle, draw, fill=blue!20,font=\bfseries}
  }
    \node[treat0] (a) at (0,0) {a};
    \node[treat1] (b) at (2,-2) {b};
    \node[treat0] (c) at (4,0) {c};
    \node[treat0] (d) at (2,2) {d};
    \node[treat0] (e) at (5,3) {e};
    
    \draw (a) -- (b);
    \draw (a) -- (d);
    \draw (b) -- (c);
    \draw (c) -- (d);
    \draw (c) -- (e);
    \draw (d) -- (e);
\end{tikzpicture}
\caption{Network  $G^\text{toy}$ with treatment assignment $\boldsymbol{t}'$.}
\label{fig:change of t}
\end{figure}

    \item We then recompute the optimality criterion, say, $\tau'$, for this modified assignment. 
    \item If $\tau'> \tau$, the updated treatment assignment $\boldsymbol{t}'$ is retained; otherwise, we revert to the previous configuration $\boldsymbol{t}$.  
\end{itemize}

This procedure is repeated iteratively for all 4 remaining nodes.
To enhance robustness, this optimization process is performed across multiple (R) randomly initialized treatment assignments.
The final treatment allocation corresponds to the configuration that yields the highest observed value of the objective function, denoted as $\tau^*$. We summarize the entire procedure in Algorithm~\ref{DOE_alg}. We can extend the Algorithm~\ref{DOE_alg} for multiple treatments in a similar manner.

\begin{algorithm*}[!ht]
\caption{\enskip Optimal Treatment Assignment}
  \label{DOE_alg}
\begin{enumerate}
 \item \textbf{Input:} Network $G$ with $n$ nodes, two treatments 
 $\{t_{(0)},t_{(1)}\}$,  number of random initializations $R$.

 \item \textbf{Output:} Optimal treatment assignment $\boldsymbol{t}^*$ and optimal criterion value $\tau^*$.
 
 \item For $r = 1$ to $R$:
  \begin{enumerate}
   \item Generate an \textit{initial} random treatment assignment vector $\boldsymbol{t}=({t}^1_{(0)}, t^2_{(1)},\ldots, t^n_{(0)})$ for the graph $G$, such that each node receives exactly one treatment from $\{t_{(0)},t_{(1)}\}$;

  \item Compute the optimality criterion $\tau$ (Using equation~\ref{Eq:OptiCr}) for the initial treatment assignment;

  \item For $i = 1$ to $n$:
    \begin{enumerate}
              \item Exchange the treatment of node $i$ with a different treatment 
              (say, $t_{(0)}^{i} \rightarrow  t_{(1)}^{i}$);  
              \item Obtain new treatment assignment vector $\boldsymbol{t}'$;
  
              \item  Compute criterion $\tau^i$ for the modified design (Using equation~\ref{Eq:OptiCr});

             \item If {$\tau^i > \tau$} Update: $\boldsymbol{t}\leftarrow\boldsymbol{t}'$,  $\tau \leftarrow \tau^i$.
    \end{enumerate}
 \end{enumerate}

\item Store: $\tau[r] \leftarrow \tau$,  
             $\boldsymbol{t}[r] \leftarrow \boldsymbol{t}$.

\item $\tau^* = \max(\tau[r])$; 
$\boldsymbol{t}^*=\boldsymbol{t}$ for corresponding $\tau^*$.

\item \textbf{return} $\tau^*$ and $\boldsymbol{t^*}$.

\end{enumerate}

\end{algorithm*}


\section{Evaluation of Causal Effects}
\label{Sec:Treatment effects}
This section evaluates the causal effects induced by the optimal treatment assignment obtained from the proposed framework in Section~\ref{Sec:Methodology}. We particularly focus on commonly used causal estimands in network experiments, including the \textit{total treatment effect} (TTE), the \textit{average treatment effect} (ATE), and the \textit{average interference effect} (AIE). 

\subsection{Causal Treatment Effects}

We begin with the linear model in Eq.~\ref{Eq:linear} introduced in Section~\ref{Sec:Model1}, where the model parameters are assumed to be identical across all nodes, representing a homogeneous setting. Although Eq.~\ref{Eq:linear} does not account for network effects, our optimality criterion incorporates the network structure with the treatment-treatment edge incidence matrix $\boldsymbol{L}$. To compute the causal treatment effects, we now extend Eq.~\ref{Eq:linear} to a heterogeneous linear outcome model, allowing each node to exhibit distinct responses to its own treatment as well as to the treatments received by its neighbors. This formulation introduces node-specific heterogeneity and effectively captures interference effects arising from the underlying network structure.

We consider two treatments $\{t_{(0)},t_{(1)}\}$, where $t_{(0)}$ is the treatment for the \textit{control group}, and $t_{(1)}$ is the treatment for the \textit{treatment group}, i.e., $t_{(0)}$ represents no treatment and $t_{(1)}$ represents treatment. In this setting, let $\boldsymbol{t}^\ast$ be the optimal treatment assignment obtained from the Algorithm~\ref{DOE_alg}. For each node $i$, let $Y_i(\boldsymbol{t}^\ast)$ denote the potential outcome under the optimal assignment $\boldsymbol{t}^\ast$. The observed outcome for node $i$ is then modeled as

\begin{equation}\label{Eq:hetero}
Y_i(\boldsymbol{t}^\ast) = \alpha_i + \beta_i t^\ast(i) + \sum_{k \in [n]} \gamma_{i(k)} t^\ast(k) +\epsilon_i,
\end{equation}
where $\alpha_i$ represents the baseline outcome for node $i$, $\beta_i$ denotes the node-specific direct treatment effect (allowing for heterogeneity across nodes), and $\gamma_{i(k)}$ captures the interference effect between node $k$ and node $i$. The error term $\epsilon_i \sim \mathcal{N}(0, \sigma^2)$ represents the random noise.
To capture variability across nodes and their connections, we assume the model parameters are drawn from normal distributions as
$\alpha_i \sim \mathcal{N}(\mu_{\alpha},\sigma_{\alpha}^2)$, 
$\beta_i \sim \mathcal{N}(\mu_{\beta}, \sigma_{\beta}^2)$, and $\gamma_{ki} \sim \mathcal{N}(\mu_{\gamma}, \sigma_{\gamma}^2)$~\citep{hu2022, Christina_2022}.

The TTE is computed by comparing outcomes between the control and treatment groups.
Formally, TTE is defined as

\begin{equation}\label{Eq:tte}
 \mathrm{TTE} = \frac{1}{n} \sum_{i=1}^n ( Y_i(\boldsymbol{1}
)-Y_i(\boldsymbol{0})) ,  
\end{equation}

where $\boldsymbol{1}$ denotes the treatment vector in which all nodes receive the treatment, and $\boldsymbol{0}$ indicates that all nodes are in the control group. This estimand captures the overall impact of universal treatment deployment~\citep{Toulis2013}.
The ATE quantifies the expected change in an outcome of a node when it alone receives the treatment, while all other units remain untreated. This estimand isolates the direct effect of the treatment on a node, excluding interference from neighbors. ATE is defined as
\begin{equation}\label{Eq:ate}
    \mathrm{ATE} =  \frac{1}{n} \sum_{i=1}^n ( Y_i(\boldsymbol{e_i}
)-Y_i(\boldsymbol{0})) ,
\end{equation}

where $\boldsymbol{e_i}$ represents the standard basis vector, which takes the value 1 at the $i^{th}$ position when node $i$ is treated, and 0 elsewhere, and $\boldsymbol{0}$ represents the all-control assignment.
The AIE captures the network effect resulting from the treatments of the neighbors. We can define the AIE as
\begin{equation}\label{Eq:aie}
    \mathrm{AIE} = \frac{1}{n} \sum_{i=1}^n ( Y_i(e_{[n]\setminus \{i\}})-Y_i(\boldsymbol{0})) ,
\end{equation}

where $e_{[n] \setminus \{i\}}$ denotes the treatment vector in which all nodes except node $i$ are treated, and $\boldsymbol{0}$ represents the all-control assignment.

Having defined the causal estimands TTE, ATE, and AIE in Eqs.~\ref{Eq:tte}, \ref{Eq:ate}, and \ref{Eq:aie}, we next describe their estimation in the presence of network interference, following the approach of \cite{Christina_2022}. Estimation is carried out under the optimal treatment assignment $\boldsymbol{t}^\ast$ obtained in Section~\ref{Sec:Optimal_Design} via Algorithm~\ref{DOE_alg}, and is based on a heterogeneous linear network effects model. The estimator for the TTE can be written as

\begin{equation}\label{Eq:tte_est}
\widehat{\mathrm{TTE}} = \frac{1}{p} \left( \frac{1}{n} \sum_{i=1}^{n} Y_i(\boldsymbol{t}^\ast) - \frac{1}{n} \sum_{i=1}^{n} \alpha_i \right) ,
\end{equation}
where $\mathbb{E}[t^\ast(i)] = p$ for all $i \in n$. The ATE estimator is formally, expressed as
\begin{equation}\label{Eq:ate_est}
    \widehat{\mathrm{ATE}} = \frac{1}{n} \sum_{i=1}^{n}\left({\frac{t^\ast(i)}{\mathbb{E}[t^\ast(i)]}- \frac{(1-t^\ast(i))\rho_i}{\mathbb{E}[t^\ast(i)](1-\rho_i)}}\right) (Y_i(\boldsymbol{t}^\ast)-\alpha_i),
\end{equation}

where the neighborhood exposure $\rho_i = \mathbb{P}(t^\ast(i)=1 \mid t^\ast(k) = 1)$ for all $(k,i)$. Finally, the AIE estimator 
is defined as

\begin{equation}\label{Eq:aie_est}
    \widehat{\mathrm{AIE}}= \frac{1}{n} \sum_{i=1}^{n} \frac{1-t^\ast(i)}{\rho_i \mathbb{E}[1-t^\ast(i)]} \left( Y_i(\boldsymbol{t}^\ast)-\alpha_i \right),
\end{equation}

where $\rho_i = \mathbb{P}(t^\ast(k)=1 \mid t^\ast(i) =0)$ for all $(k,i)$.

\subsection{Properties of Causal Treatment Effect Estimators}
To evaluate the estimation error and uncertainty of the TTE, ATE, and AIE estimators, we compute the bias and standard error (SE) for each estimator across B bootstrap network simulations. Let $\hat{\theta}_1, \hat{\theta}_2, \dots, \hat{\theta}_B$ denote the estimated values obtained from Eq.~\ref{Eq:tte_est},~\ref{Eq:ate_est},~\ref{Eq:aie_est} and let $\theta_1, \theta_2, \ldots\theta_B$ denotes the true estimand obtained form the Eq.~\ref{Eq:tte},~\ref{Eq:ate},~\ref{Eq:aie} over $B$ simulations. The bias estimate is then given by the mean bias for the B simulations as 
\begin{equation}\label{Eq:bias}
\widehat{\text{Bias}} = \frac{1}{B} \sum_{b=1}^B (\hat{\theta}_{b} - \theta_{b}). 
\end{equation}
The standard error of the bias quantifies the variability of the estimator across simulations and is computed as
$\text{SE} = \frac{S}{\sqrt{B}}$, where  \resizebox{0.35\linewidth}{!}{$S^2 = \frac{1}{B-1} \sum_{b=1}^B \left((\hat{\theta}_{b} -\theta_{b})- \widehat{\text{Bias}}\right)^2$}.



\section{Simulation Studies}
\label{Sec:Sim studies}
In this section, we conduct extensive simulation studies to demonstrate how our proposed optimal design framework works on random networks with different network geometries. 

\subsection{Random Networks}
\label{Subsec:Random Net}
We apply the algorithm to four network models, i.e., the Erd\H{o}s--R\'enyi Graph, geometric random graph, preferential attachment model, and stochastic block model. Below, we define the network models used in our analysis. 

\begin{itemize}
\item
\textbf{Erd\H{o}s--R\'enyi graph:} The Erd\H{o}s--R\'enyi (ER) graph, denoted by $G(n,p) = (V, E)$, is a random graph with $n$ nodes, where each possible edge $(i,j)$, for $i \neq j$, is independently included in the edge set $E$ with probability $p$.
The resulting graph structure $G = (V,E)$ exhibits key statistical properties: the expected number of edges is $\mathbb{E}(|E|) = \binom{n}{2}p$ and the degree of each node follows a Binomial distribution, $\mathrm{Binomial}(n-1, p)$~\citep{Erdos_1959, Erdos_1960, Newman_2002}.

\item
\textbf{Geometric Random Graph:} The Geometric Random (GR) graph, denoted by $G(n,r)$, incorporates spatial constraints into network formation, making it particularly suitable for modeling cyber-physical and spatial systems such as wireless communication, sensor, and transportation networks. The GR graph $G(n,r) =(V,E)$ is a graph, where $n$ nodes are randomly positioned within a geometric space, and an edge $(i,j)$ is formed if the Euclidean distance between nodes $i$ and $j$ is less than a specified connection radius $r$. The expected degree of a node in a GR graph depends on both the connection radius $r$ and the node density, which together determine the overall connectivity of the network~\citep{Dall_2002, penrose2003}.

\item  
\textbf{Preferential Attachment Model:} The Preferential Attachment (PA) model generates networks with a power-law degree distribution~\citep{newman2005power}, a pattern often seen in real systems such as the internet, citation networks, and social graphs. The model builds a network by adding nodes one at a time. Each new node connects to a set of ($m$) existing nodes, with higher-degree nodes having a greater chance of being chosen. This ``rich get richer'' dynamic makes well-connected nodes increasingly likely to attract additional links~\citep{Barabasi_1999}.

\item \textbf{Stochastic Block Model:} The Stochastic Block Model (SBM) is a generative model for random graphs with community structure. In an SBM $G$, the node set $V$ is partitioned into $K$ disjoint communities, and the probability of an edge between two nodes depends on their community memberships. Formally, it is a graph with $n$ nodes, where edges are independently generated according to two probabilities: $p_w$ for edges within the community and $p_b$ for edges between communities.
When $p_w > p_b$, nodes are more likely to connect within their own communities than across different ones, resulting in clustered network patterns. In the balanced SBM, communities are of equal size, whereas in the unbalanced SBM, community sizes vary across groups. Such structures are commonly observed in social and biological systems. The expected degree of each node depends on its community size and the connection probabilities $p_w$ and $p_b$~\citep{HOLLAND_1983,Emmanuel_2018}.

\end{itemize}

Table~\ref{Tab:RG properties} summarizes the key structural characteristics of the networks used in this study, and Supplementary Material Figure~S1 and S2 illustrate their corresponding network structure.

\begin{table*}[!ht]
	\caption{Properties of the four random networks.}
	\centering
	\begin{tabular}{l*{6}{c}r} \hline
		Network & Nodes & Parameters & 	Edges \\
		\hline
		ER graph & 250 & $p$ = 0.02 & 583\\
        GR graph & 250 & $r$ = 0.1 & 910\\		
		PA model & 250 & $m$ = 1 & 249\\
        SBM-balanced & 250 & $K$ = 5, $p_{w}= 0.3$, $p_{b}=0.01$, $n = (50,50,50,50,50)$ & 2137\\
        SBM-unbalanced & 250 & $K$ = 5, $p_{w}= 0.3$, $p_{b}=0.01$, $n = (90,60,45,35,20)$  & 2541\\
		\hline
	\end{tabular}
    \label{Tab:RG properties}
\end{table*}

\newpage
\subsection{Contenders and Baseline}

We compare our proposed optimal design framework with existing methods for treatment allocation. Specifically, we consider the following competing approaches:

\begin{itemize}
    \item \textit{Randomized Design}: Randomized treatment assignment consists of independently assigning each node to treatment, i.e., treatment assignments are independent across nodes~\citep{EcklesKarrerUgander_2017}.
    In this setting, the treatment indicator for each node is modeled as an independent and identically distributed Bernoulli random variable,
    \begin{equation*}
        t(i) \sim \mathrm{Bernoulli}(p) ,      i = 1, \cdots n,
    \end{equation*}

where $p$ is the probability of treatment assignment.

    \item \textit{Graph Cluster Randomization}: In standard graph cluster randomization~\citep{ugander2013}, nodes are partitioned into $N_C$ clusters $C_1, C_2, \cdots C_{N_C}$ and treatment is assigned at the cluster level. Specifically, each cluster $C_d$ is independently assigned a treatment 
    \begin{equation*}
        W_d \sim \mathrm{Bernoulli}(p) , \quad d = 1, \dots, N_C.
    \end{equation*}
    The treatment received by node $i$ is then given by
    \begin{equation*}
        t(i)= W_{C(i)},
    \end{equation*}
where $C(i) \in {1,\dots,N_C}$ denotes the index of the cluster containing node $i$.
This approach incorporates network structure by assigning the same treatment to densely connected groups of nodes.

    \item \textit{Rerandomization}: Rerandomization is a design technique in which treatment assignments are repeatedly generated and accepted only if they satisfy prespecified conditions on design statistics determined by the network structure~\citep{Zhang2025}. 
    These constraints are used to improve balance and network-based characteristics of the assignment relative to purely random designs.
\end{itemize}

All designs are evaluated using the same design criterion $\tau$ (Eq. \ref{Eq:OptiCr}), ensuring a consistent basis for comparison. The performance of the proposed optimal design is evaluated against the network models described in Section~\ref{Subsec:Random Net}. 
In addition, we compare the treatment balance across different treatment allocation methods. The results are presented in the following sections.

\subsection{Experiments on ER, GR, and the PA models}

In this experiment, we generate ER, GR, and the PA networks with properties mentioned in Table~\ref{Tab:RG properties}. For these three networks, we evaluate criterion values and treatment balances for the proposed optimal design and its competitors. 

\begin{figure*}[!ht]
     \centering
     \begin{tabular}{cc}
      \includegraphics[width=0.4\textwidth] {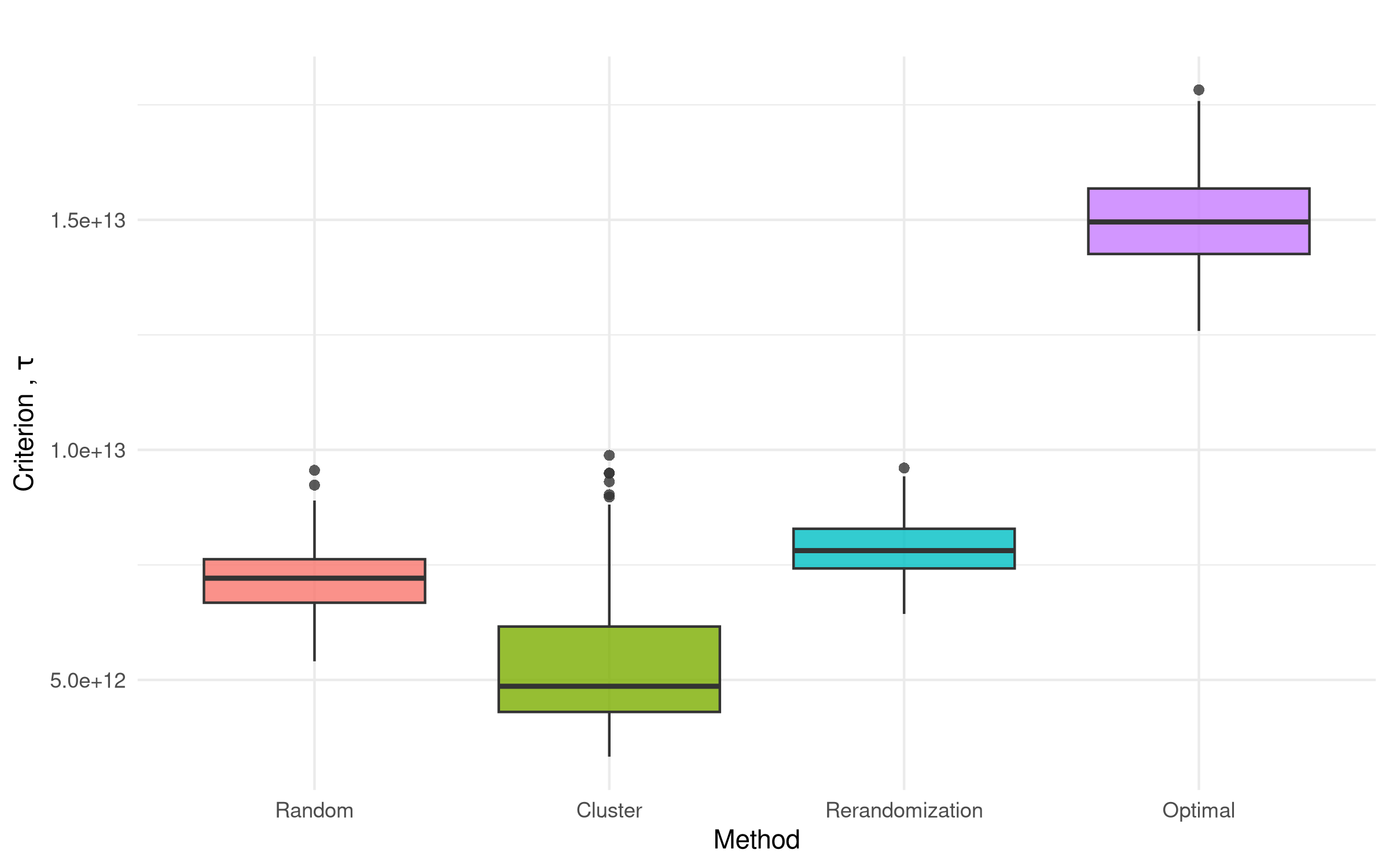} &
      \includegraphics[width=0.40\textwidth]{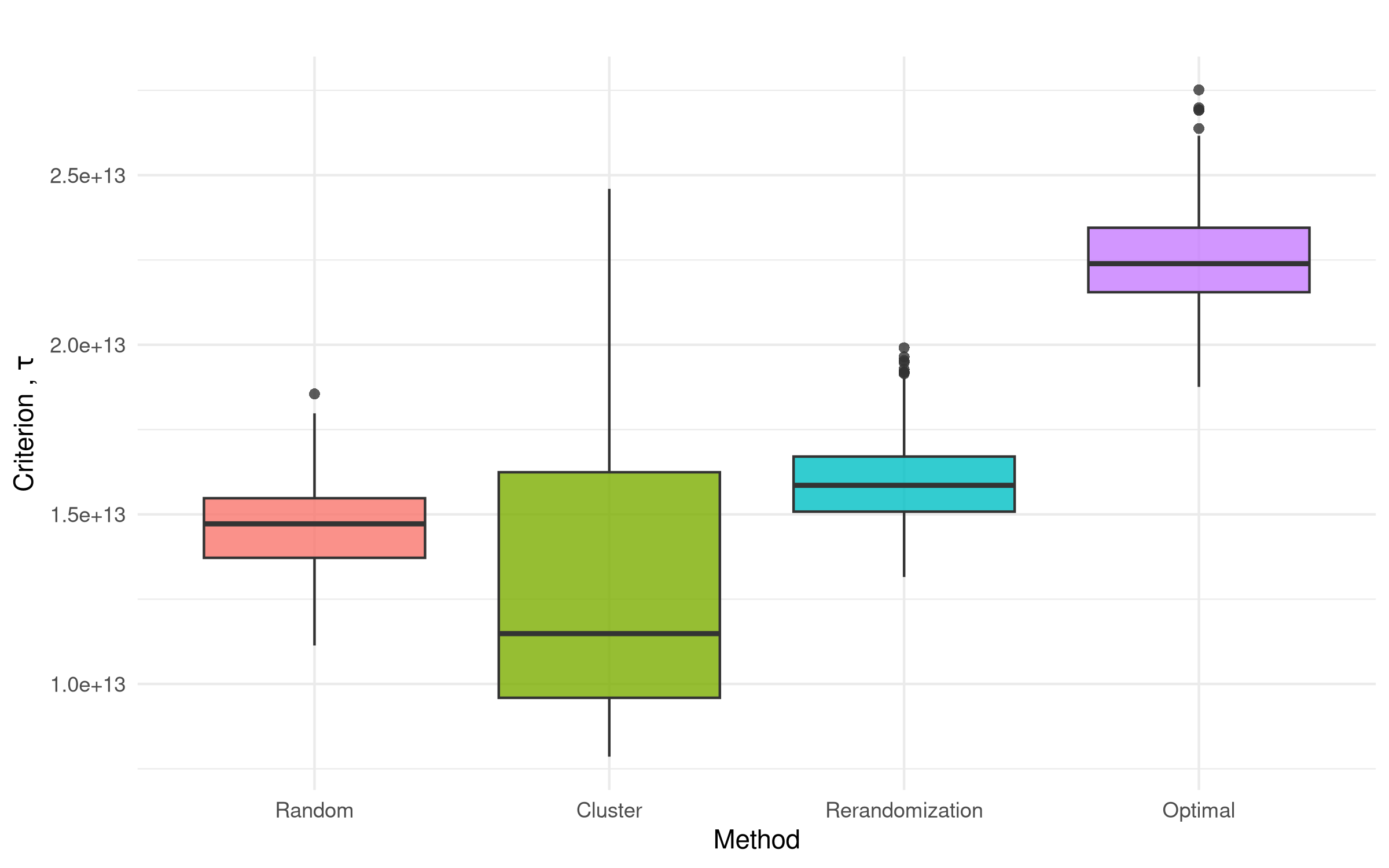} \\
       (a) ER graph. & (b) GR graph. \\
    \multicolumn{2}{c}{
     \includegraphics[width=0.45\textwidth]{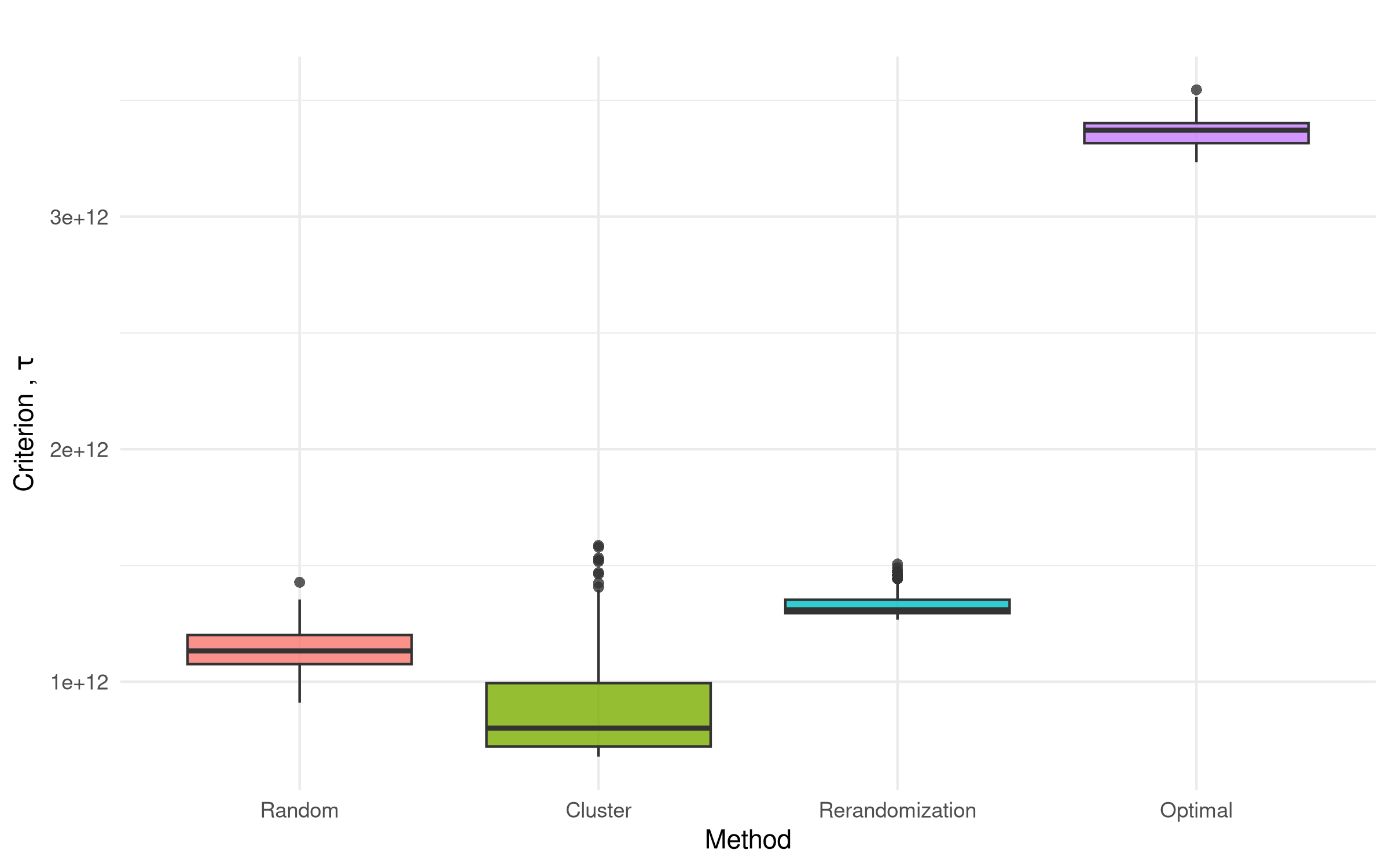}
     }\\ 
      \multicolumn{2}{c}{(c) PA model.} \\
      \end{tabular}
	\caption{Design criteria for different random networks.}
        \label{fig:Comparison}
\end{figure*}

Figure~\ref{fig:Comparison} compares the criterion values obtained by the four methods. Across all three network settings, the proposed optimal design achieves the highest criterion values, indicating that the local search algorithm consistently improves the Fisher information–based design criterion relative to the benchmark methods. Among the benchmarks, rerandomization generally performs best, followed by Bernoulli randomization, while cluster randomization exhibits lower and more variable performance. The consistency of this ordering across the three network models suggests that the proposed design is robust to differences in network topology.

\begin{figure*}[!ht]
     \centering
     \begin{tabular}{cc}
      \includegraphics[width=0.35\textwidth]{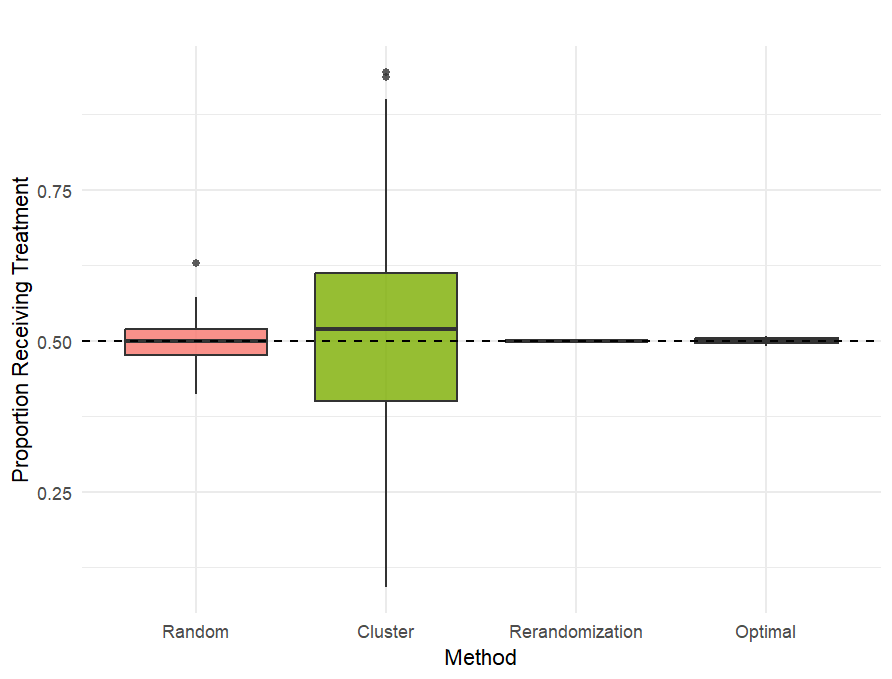} &
      \includegraphics[width=0.35\textwidth]{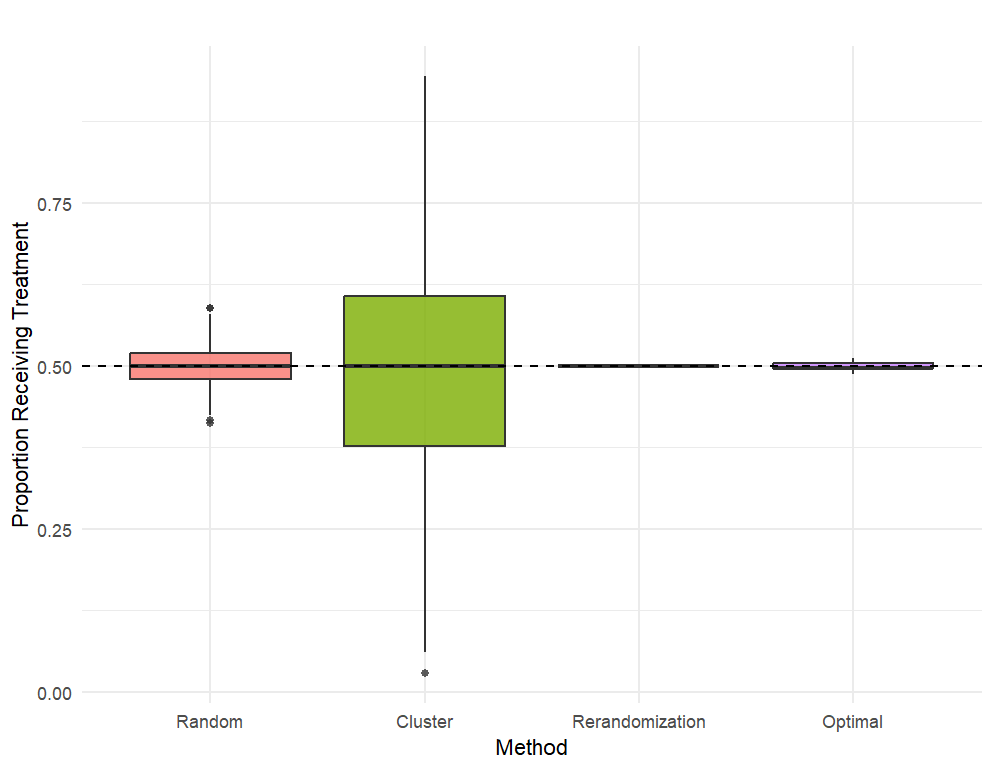} \\
       (a) ER graph. & (b) GR graph. \\
    \multicolumn{2}{c}{
     \includegraphics[width=0.40\textwidth]{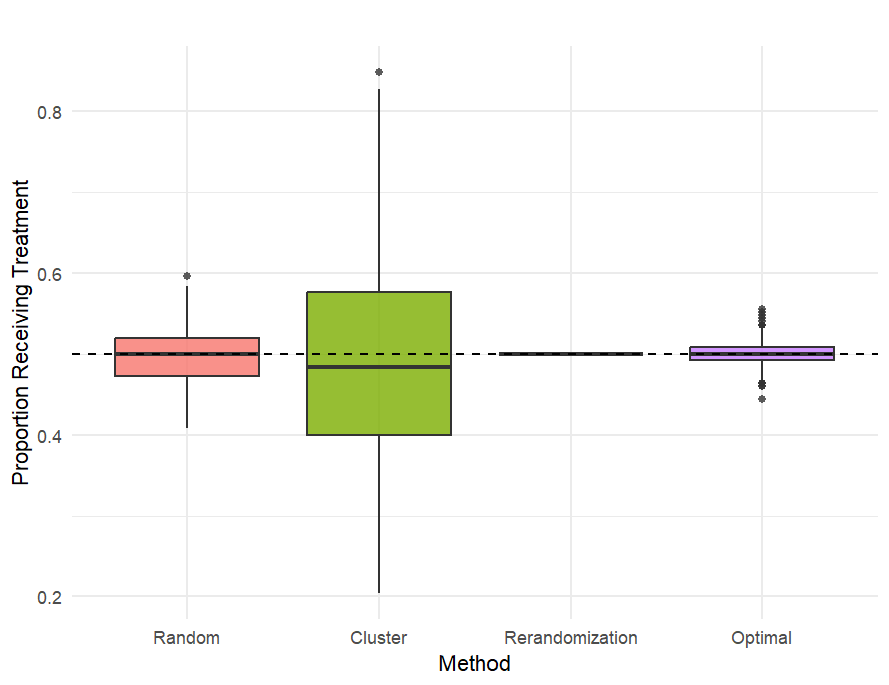}}\\ 
      \multicolumn{2}{c}{(c) PA model.} \\
      \end{tabular}
	\caption{Treatment balance for different random networks.}
        \label{fig:Treatment balance}
\end{figure*}

Figure~\ref{fig:Treatment balance} summarizes the proportion of nodes assigned to treatment under the four assignment methods. 
The proposed optimal design is similarly concentrated around 0.5, indicating that the optimization procedure maintains strong treatment balance. Rerandomization achieves exact balance, as its assignment mechanism yields equal numbers of treated and control nodes. By contrast, Bernoulli randomization is balanced only in expectation and consequently exhibits moderate variability across runs, while cluster randomization displays the greatest instability, with some realizations leading to substantial imbalance.

\newpage
\subsection{Experiment on SBMs}
\begin{figure*}[!ht]
     \centering
     \begin{tabular}{cc}
      \includegraphics[width=0.40\textwidth]{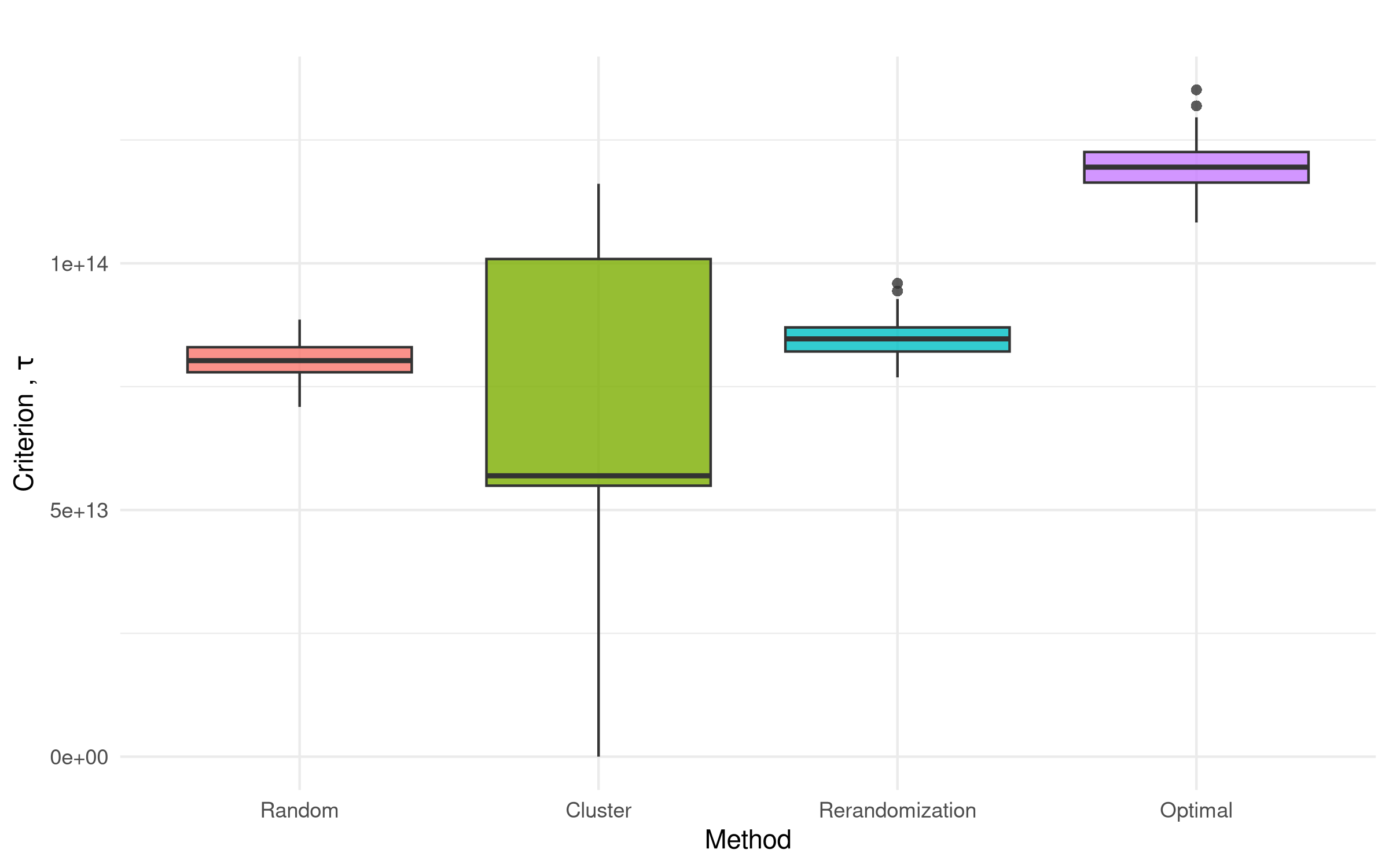} &
      \includegraphics[width=0.40\textwidth]{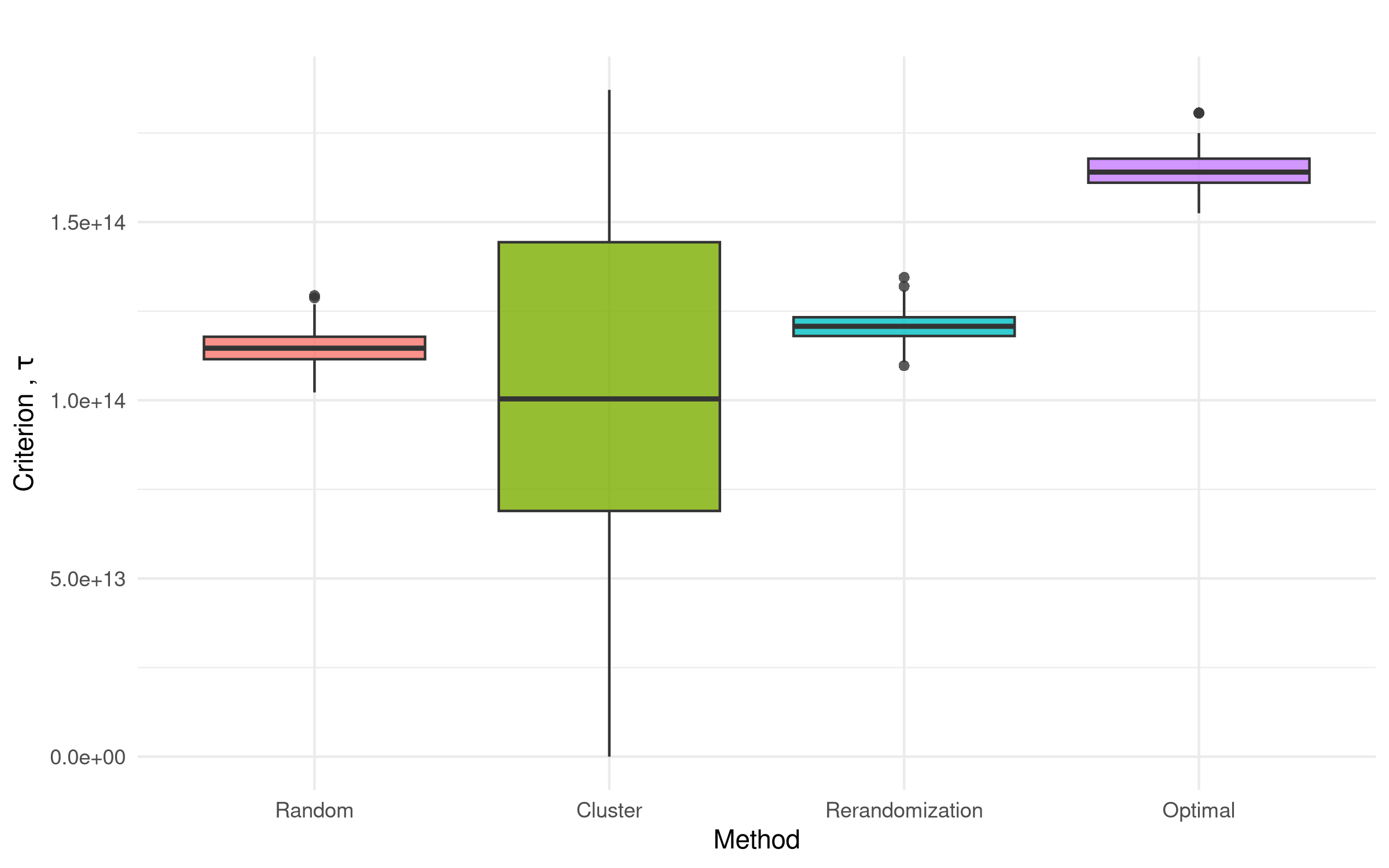} \\
       (a) Balanced. & (b) Unbalanced. \\
      \end{tabular}
	\caption{Design criteria for SBMs.}
        \label{fig:Comparison SBM}
\end{figure*}

We now perform the second experiment on balanced and unbalanced SBMs. Figure~\ref{fig:Comparison SBM} compares the criterion values across assignment methods for both balanced and unbalanced SBMs. In both settings, the proposed optimal design consistently attains the largest criterion values with lower variances.
The ranking of methods is preserved across the two SBM specifications, indicating that the proposed method remains robust to community-size imbalance. Cluster randomization exhibits the greatest variability, particularly in the unbalanced SBM, reflecting the instability induced by cluster-level assignment under uneven block structures. 

\begin{figure*}[!ht]
     \centering
     \begin{tabular}{cc}
      \includegraphics[width=0.39\textwidth]{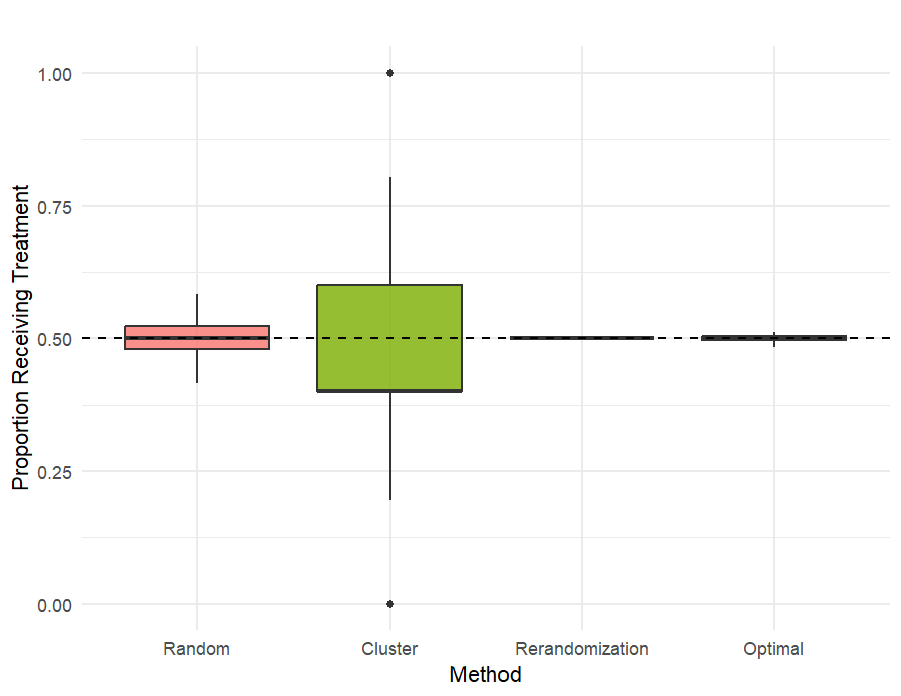} &
      \includegraphics[width=0.465\textwidth]{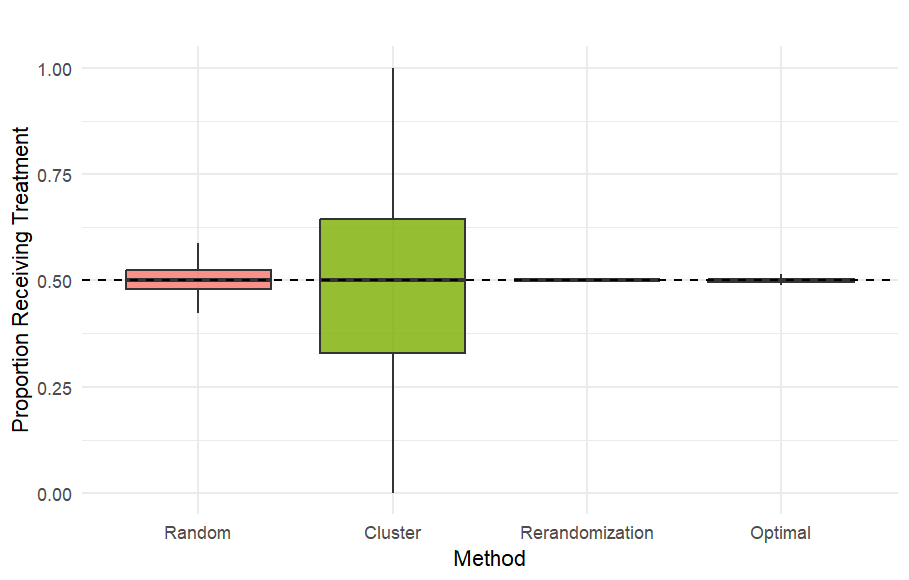} \\
       (a) Balanced. & (b) Unbalanced. \\
      \end{tabular}
	\caption{Treatment balance for SBMs.}
        \label{fig:treatment balance for SBMs}
\end{figure*}

Figure~\ref{fig:treatment balance for SBMs} shows that the proposed optimal design maintains treatment allocations close to 0.5 under both balanced and unbalanced SBM structures. This is comparable to rerandomization, which is balanced by construction, and contrasts with Bernoulli randomization and especially cluster randomization, both of which display greater variability in realized treatment proportions. The pattern is qualitatively unchanged across the two SBM specifications, suggesting that the balance properties of the proposed design are robust to block-size imbalance.

\subsection{Treatment-Control Experiment using Proposed Optimal Design}

In this experiment, for each random network, we consider two treatments $\{t_{(0)}, t_{(1)}\}$, where $t_{(0)}$ represents treatment for the control group, and $t_{(1)}$ represents treatment for the treatment group. Each node in the network is assigned to either the control group or the treatment group. To determine the optimal treatment assignment across nodes, we apply our proposed optimal design framework, Algorithm~\ref{DOE_alg}, with $R=20$, i.e., we perform 20 distinct initial treatment assignments. The model parameters of Eq.~\ref{Eq:hetero} are drawn from normal distributions as follows $\epsilon_i \sim \mathcal{N}(0, 1)$, $\alpha_i \sim \mathcal{N}(1,0.25)$, 
$\beta_i \sim \mathcal{N}(5, 0.25)$, and $\gamma_{ki} \sim \mathcal{N}(5, 0.25)$.

\begin{figure*}[!ht]
     \centering
     \begin{tabular}{cc}
      \includegraphics[width=0.39\textwidth] {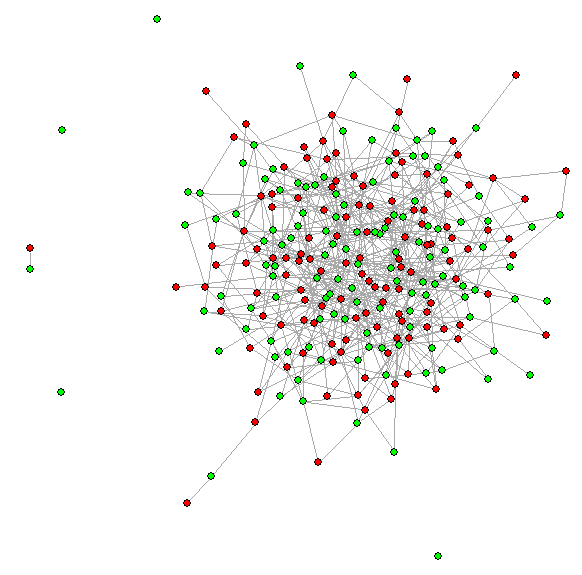} &
      \includegraphics[width=0.31\textwidth]{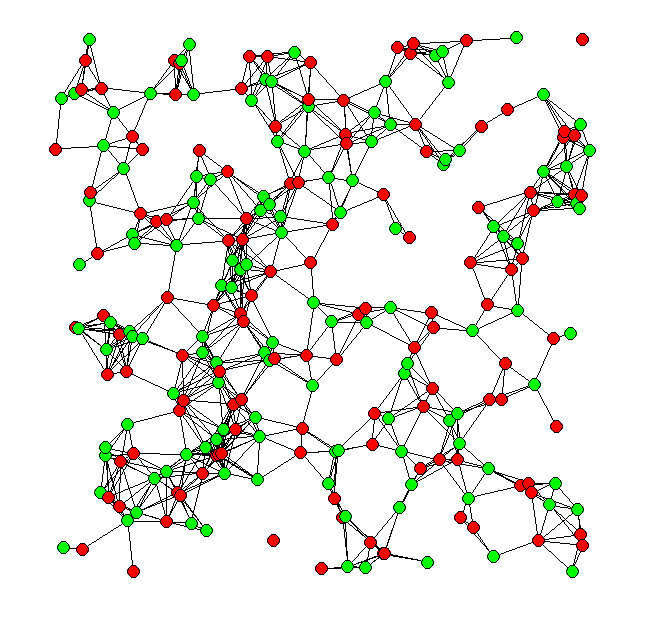} \\
       (a) ER graph. & (b) GR graph. \\
    \multicolumn{2}{c}{
     \includegraphics[width=0.32\textwidth]{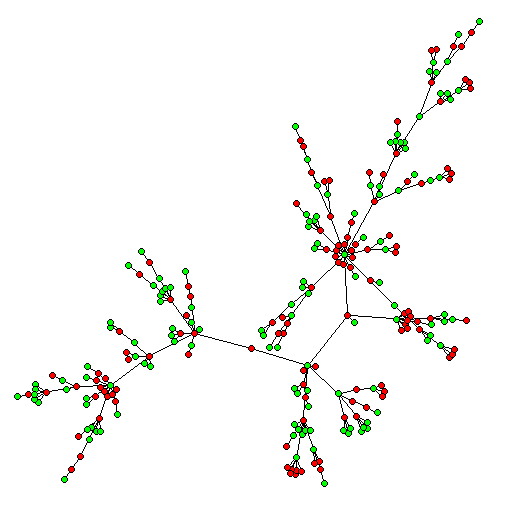}
     }\\ 
      \multicolumn{2}{c}{(c) PA model.} \\
      \end{tabular}
	\caption{Optimal treatment assignment $\boldsymbol{t}^\ast$ for different random networks. Nodes shown in red receive treatment $t_{(0)}$, while nodes shown in green receive treatment $t_{(1)}$.}   
        \label{fig:Optimal t RG}
\end{figure*}

\begin{figure*}[!ht]
     \centering
     \begin{tabular}{cc}
      \includegraphics[width=0.30\textwidth]{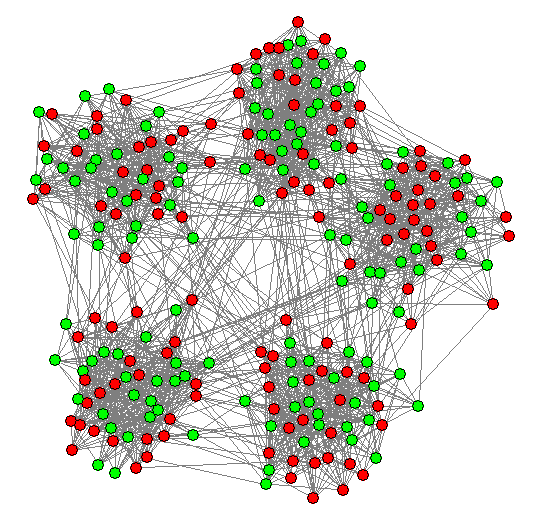} &
      \includegraphics[width=0.32\textwidth]{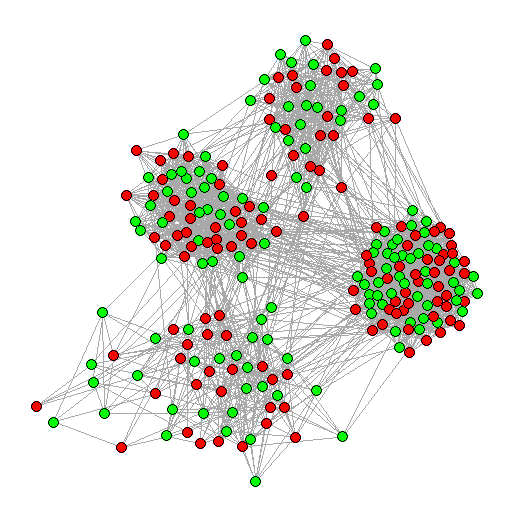} \\
       (a) Balanced. & (b) Unbalanced. \\
      \end{tabular}
	\caption{Optimal treatment assignment $\boldsymbol{t}^\ast$ for SBMs. Nodes shown in red receive treatment $t_{(0)}$, while nodes shown in green receive treatment $t_{(1)}$.}    
        \label{fig:Optimal t SBM} 
\end{figure*}

The figures~\ref{fig:Optimal t RG} and~\ref{fig:Optimal t SBM}  illustrate the resulting optimal treatment assignment $\boldsymbol{t}^\ast$ corresponding to the highest optimal criterion value $\tau^*$. The red nodes represent treatment $t_{(0)}$, and the green nodes represent treatment $t_{(1)}$. 
We find that across the four network models, the optimal assignment of treatment $t_{(0)}$ (red) and treatment $t_{(1)}$ (green) follows patterns shaped by each network's geometry. 
In the ER graph, the two assignments are well mixed, 
consistent with the absence of strong structural features. 
The algorithm distributes $t_{(0)}$s and $t_{(1)}$s to balance exposure across a largely homogeneous set of nodes.
In the GR graph, spatial clustering leads to visible pockets where treatment $t_{(1)}$ gathers, while treatment $t_{(0)}$ appears more frequently in sparse regions or boundary areas. 
The PA model highlights the influence of hubs, i.e., high-degree nodes often share the same assignment, and long branches exhibit stretches of consistent labels that follow the network’s tree-like shape. In both balanced and unbalanced SBM, community structure drives the clearest pattern, with some blocks receiving a heavier share of $t_{(1)}$ and others leaning toward treatment $t_{(0)}$. 


We now compute the causal estimators, i.e., TTE, ATE, and AIE, 
under each network model and evaluate their performance in terms of bias and standard error.
Table~\ref{tab:graph_results} summarizes the resulting biases and standard errors for $B=100$ simulations for all five network settings.

\begin{table}[ht!]
\centering
\caption{Estimation results for different graph models}
\label{tab:graph_results}

\begin{tabular}{c|c|c|c}
\hline
Graph Model & Treatment Effect & Bias & Standard Error\\ 
\hline

 & TTE & -0.077& 0.073\\
ER graph & ATE & 5.226 & 0.090 \\
 & AIE & -848.3 & 618.78\\
\hline

 & TTE & 0.019 & 0.086\\
GR graph & ATE & 12.214 & 0.127 \\
 & AIE & 431.402 & 566.15\\
\hline

 & TTE & 0.066 & 0.059\\
PA model & ATE & 0.584 & 0.037\\
 & AIE & 0.014 & 0.062\\
\hline

  & TTE & 0.106 & 0.129\\
SBM-balanced & ATE & 29.508 & 0.165\\
 & AIE & -0.162 & 0.134\\
\hline

  & TTE & -0.066 & 0.129 \\
SBM-unbalanced & ATE & 36.535 & 0.182\\
 & AIE & 0.057 & 0.135\\
\hline
\end{tabular}%
\end{table}

The differences in estimator performance across different graph models arise from how each network structure influences the distribution of the neighborhood exposure $\rho_i$ under the optimized treatment assignment.
The ATE estimator depends on exposure among treated nodes, i.e., nodes that receive treatment $t_{(1)}$. When the optimized assignment induces substantial variability in $\rho_i$ across treated units, the resulting ATE bias increases. 
This pattern is reflected in the ER, GR, and especially in the balanced and unbalanced SBM graphs, where the observed ATE biases indicate differing levels of heterogeneity in $\rho_i$ among treated nodes.
By contrast, the AIE estimator depends on exposure among control nodes that receive treatment $t_{(0)}$. In this case, even a small number of control nodes with very low exposure can cause instability in the estimator because the AIE estimator incorporates the factor $1/{\rho_i}$. This behavior appears in the ER and GR graphs, where the large AIE biases suggest that the optimized design produced some control nodes with near-zero exposure. In the SBMs and PA model, however, the small AIE biases imply that the optimized assignments generated more stable exposure levels among control nodes in these networks.
The comparatively small ATE and AIE biases in the PA model suggest that its hub-dominated structure leads to more uniform exposure patterns for both treated and control nodes under the optimized assignment.
Across all network models, the TTE estimator remains stable because it does not depend on neighborhood exposure and is therefore largely unaffected by the structural differences between graphs.


\subsection{Three-Treatment Experiments using Proposed Optimal Design}

We now extend the experiment beyond a treatment-control framework. Here, we consider three active treatments, denoted as $\{t_{(0)}, t_{(1)}, t_{(2)}\}$, for example, three distinct COVID-19 intervention strategies such as vaccination, masking, and social distancing. Each node in the random network receives one of the three treatments. The goal is to find the optimal allocation of these treatments across the network's nodes using the proposed optimal design strategy. Similar to the treatment-control experiments, we chose $R=20$. However, causal estimands such as the TTE, ATE, and AIE cannot be computed in settings with three treatments, as these quantities are defined exclusively within the conventional treatment-control framework.

\begin{figure*}[!ht]
     \centering
     \begin{tabular}{cc}
      \includegraphics[width=0.40\textwidth] {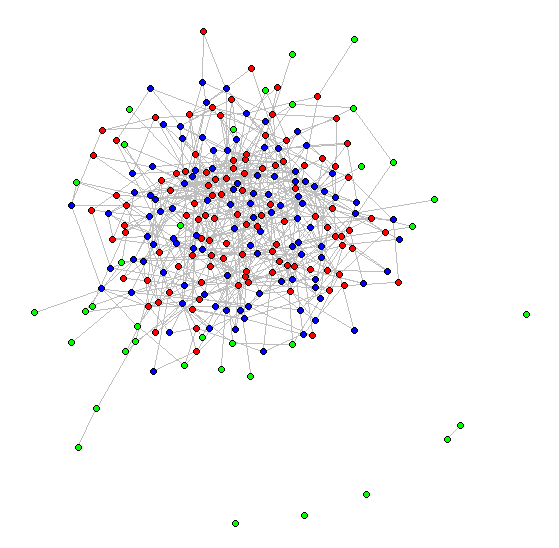} &
      \includegraphics[width=0.30\textwidth]{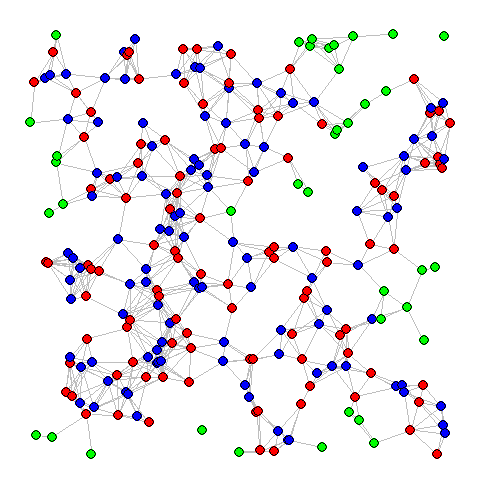} \\
       (a) ER graph. & (b) GR graph. \\
    \multicolumn{2}{c}{
     \includegraphics[width=0.34\textwidth]{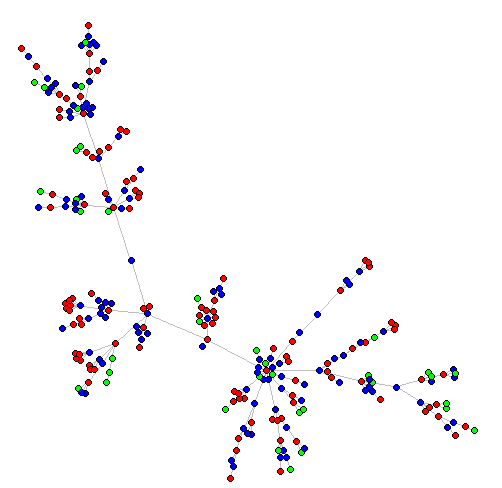}
     }\\ 
      \multicolumn{2}{c}{(c) PA model.} \\
      \end{tabular}
	\caption{Optimal treatment assignment for different random networks with three treatments.}
        \label{fig:Supp_Opt_t t=3}
\end{figure*}

\begin{figure*}[!ht]
     \centering
     \begin{tabular}{cc}
      \includegraphics[width=0.33\textwidth]{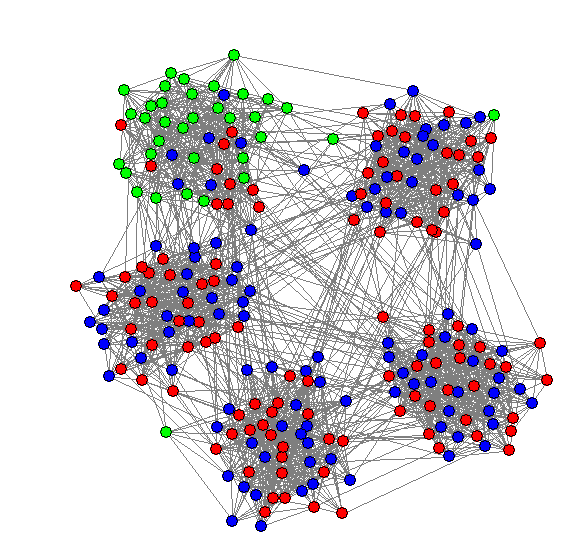} &
      \includegraphics[width=0.31\textwidth]{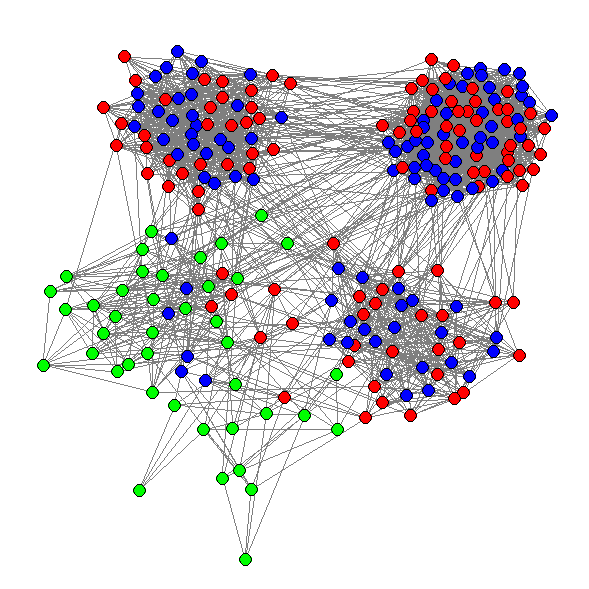} \\
       (a) Balanced. & (b) Unbalanced. \\
      \end{tabular}
	\caption{Optimal treatment assignment for SBMs with three treatments.}
        \label{fig:Supp_Opt_t t=3 SBM}
\end{figure*}

Figure~\ref{fig:Supp_Opt_t t=3} and~\ref{fig:Supp_Opt_t t=3 SBM} show the optimal treatment assignment $\boldsymbol{t}^\ast$ in the three-treatment setting, corresponding to the optimal criterion value $\tau^{\ast}$. Red, green, and blue nodes indicate treatments $t_{(0)}$, $t_{(1)}$, and $t_{(2)}$, respectively. We find that across all five network settings, the optimal assignment exhibits patterns that reflect the underlying geometry of each network.

In the ER graph, two treatments appear well mixed throughout the dense core of the network, while one treatment appears to be in the boundary regions. In the GR graph, spatial layout creates visible clusters: nodes located in dense spatial regions often share the same treatments, whereas nodes near boundaries or sparsely connected zones receive a different treatment. The PA model shows treatment patterns that follow its tree-like growth, with long branches frequently sharing the same treatments and high-degree hubs forming focal points for particular treatments. In the SBMs, community structure produces the most distinct treatment groupings, with blocks tending to concentrate specific treatments while still allowing some mixing along the boundaries where communities overlap. 

\newpage

\section{Real Data Analysis}
\label{Sec:real}
In this section, we apply the proposed optimal design framework to two real-world network datasets: a college network and an ego-Facebook network.


\subsection{College Network}

\begin{figure*}[!ht]
     \centering
     \begin{tabular}{cc}
      \includegraphics[width=0.40\textwidth]{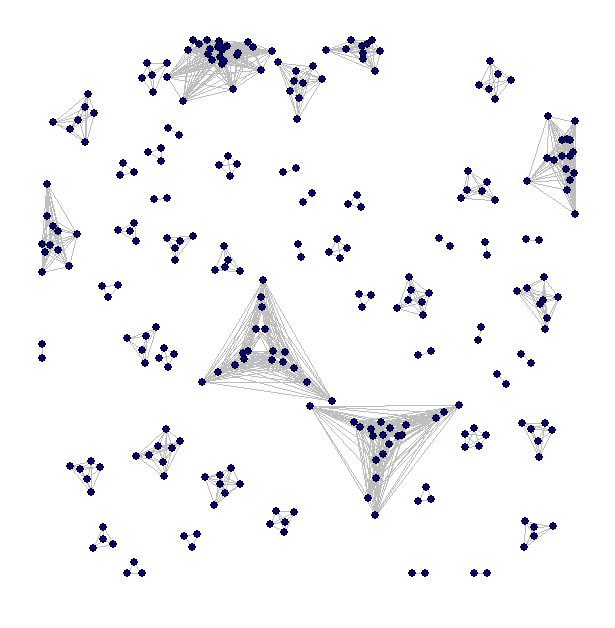} &
      \includegraphics[width=0.40\textwidth]{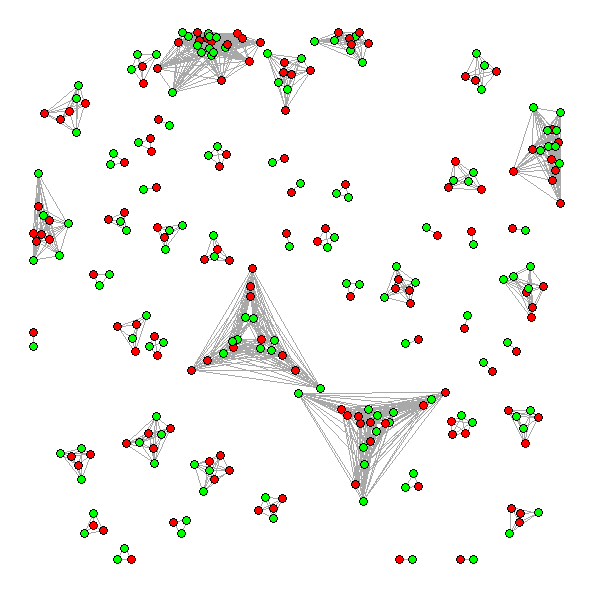} \\
       (a) College housing networks. & (b) Optimal treatment assignment. \\
      \end{tabular}
	\caption{College housing networks for the fall 2019 (a) and optimal treatment assignment under two treatments, where red and green nodes correspond to treatments $t_{(0)}$ and $t_{(1)}$ respectively (b).}
      \label{fig:College Network}
\end{figure*}

The college network dataset contains anonymized housing records for undergraduate students at the State University of New York at Geneseo for fall 2019~\citep{hartvigsen_2021}. The dataset contains residential connections and has been used to study public health interventions such as masking, testing, quarantine, and social distancing during the COVID-19 pandemic. Our study uses a network with 278 nodes (students) and 1193 edges (shared residences). Students who live alone are not included, and it is assumed that no students leave mid-semester. Figure~\ref{fig:College Network} (a) shows the resulting network, which contains several large, dense clusters along with many small groups and isolated pairs. The large clusters likely correspond to residence halls or large apartment complexes, while the smaller components reflect small housing units.

Our objective is to assess how non-pharmaceutical interventions can be strategically deployed across a college contact network to mitigate COVID-19 transmission. We begin by considering a treatment-control framework with two treatments, $t_{(0)}$ and $t_{(1)}$, which may correspond, for example, to masking and no masking. To determine an optimal allocation of treatments across the network, we apply Algorithm~\ref{DOE_alg}, which explicitly accounts for the underlying network structure and potential interference among connected individuals.

The optimal treatment distribution $\boldsymbol{t}^\ast$ corresponding to $\tau^\ast$, as shown in Figure~\ref{fig:College Network} (b), assigns treatments $t_{(0)}$ (red) to 140 nodes and $t_{(1)}$ (green) to 138 nodes across the network. The overall distribution indicates that both treatments are dispersed throughout the network, with certain dense clusters exhibiting a slight predominance of $t_{(1)}$ over $t_{(0)}$, while no subgraph is dominated by a single treatment.

Table~\ref{tab: Estimation Results for College Network} presents the estimated bias of the TTE, ATE, and AIE estimators. The network exhibits a highly heterogeneous connectivity pattern with several dense cliques embedded within a sparse structure.
Treated nodes inside dense cliques experience very high neighborhood exposure $\rho_i$, while treated nodes in sparse regions have much lower $\rho_i$.
This strong heterogeneity in neighborhood exposure among treated nodes leads to the substantial ATE bias observed in Table~\ref{tab: Estimation Results for College Network}. 
Although control nodes in dense cliques have high neighborhood exposure $\rho_i$, most control nodes are located in moderately connected or sparse regions where $\rho_i$ is low. These mixed contributions tend to offset each other, resulting in only a small AIE bias.
The TTE estimator remains consistent, as it does not depend on neighborhood exposure patterns and is therefore less affected by clustering.

\begin{table*}[!ht]
	\caption{Estimation results for the college network.}
	\label{tab: Estimation Results for College Network}	
	\centering
	\begin{tabular}{l*{6}{c}r} \hline
		Estimand & Bias \\ 
		\hline
		TTE & -0.344 \\	
		ATE & 18.558 \\
		AIE & 0.423  \\
		\hline
	\end{tabular}
\end{table*}

Now we consider three active treatments, $t_{(0)}$, $t_{(1)}$, and $t_{(2)}$, which may correspond, for example, to masking, testing, and quarantining, respectively. We apply Algorithm~\ref{DOE_alg} to obtain an optimal allocation of treatments across the network. Figure~S3 in the Supplementary Material shows the optimal treatment distribution $\boldsymbol{t}^\ast$. We find that nodes within the large, dense cliques in the network receive a mixture of two treatments, while many of the small or isolated clusters often share the same treatment.

\subsection{Ego-Facebook Network}

\begin{figure*}[!ht]
     \centering
     \begin{tabular}{cc}
      \includegraphics[width=0.40\textwidth]{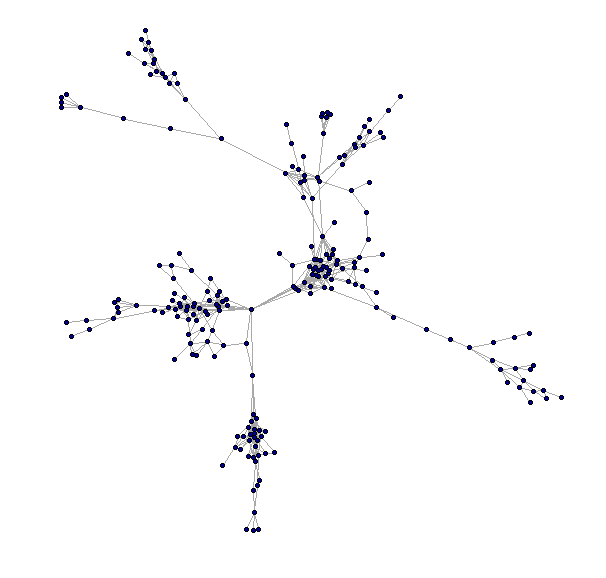} &
      \includegraphics[width=0.40\textwidth]{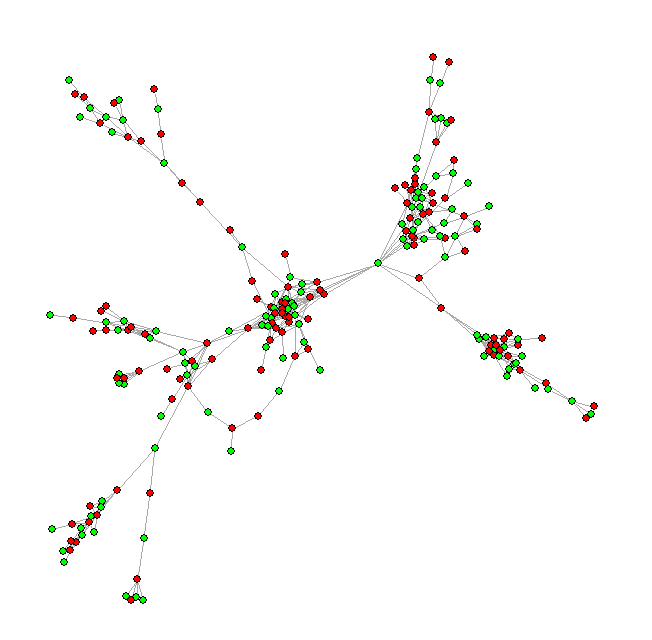} \\
       (a) Ego-Facebook network. & (b) Optimal treatment assignment. \\
      \end{tabular}
	\caption{Ego-Facebook network (a) and the optimal assignment of two treatments (b), where red and green nodes represent treatments $t_{(0)}$ and $t_{(1)}$, respectively.}
      \label{fig:Ego_facebook}
\end{figure*}

We now consider the analysis of the ego-Facebook network, which is obtained from the Stanford Network Analysis Project (SNAP)~\citep{McAuley2012}, where each node represents a single user, and the edges are ties among their friends. Figure~\ref{fig:Ego_facebook} (a) displays the ego-Facebook network used in our analysis, which contains 220 nodes and 576 edges. Although the original study examined political group affiliations, in our setting, the treatments represent hypothetical interventions assigned to nodes in the network, such as exposure to an advertisement, message, or product recommendation. Our goal is to assess how the structure of a real-world social network influences the optimal distribution of treatments under interference.

We consider a treatment-control framework with a binary treatment assignment $\{t_{(0)},t_{(1)}\}$, 
which may represent, for example, an advertisement (treatment) and no advertisement (control). The optimal treatment assignment $\boldsymbol{t}^\ast$ corresponding to $\tau^*$ is presented in Figure~\ref{fig:Ego_facebook} (b) and allocates treatment $t_{(0)}$ (red) to 108 nodes and treatment $t_{(1)}$ (green) to 112 nodes. Overall, both $t_{(0)}$ and  $t_{(1)}$ are distributed across the network. The dense cliques contain a mixture of treatments, whereas long branches and peripheral nodes tend to exhibit more homogeneous treatment assignments.


Table~\ref{tab: Estimation Results for Ego} summarizes the bias of the three causal estimands, TTE, ATE, and AIE,  under the optimal treatment assignment $\boldsymbol{t}^\ast$, providing a comparison of how each estimator varies under the interference structure of the ego-Facebook network. 
The estimated treatment effects for the ego-Facebook network reveal clear differences across the three causal treatment effects. The ego-Facebook network has a sparse, locally clustered structure composed of several mini-hubs connected by long chains of low-degree nodes. Treated nodes located within the central clusters experience very high neighborhood exposure $\rho_i$, while treated nodes along the long peripheral branches have low $\rho_i$. This heterogeneity in $\rho_i$s among treated nodes drives the ATE bias reported in Table~\ref{tab: Estimation Results for Ego}.
Although some control nodes within the clusters have high $\rho_i$s, many control nodes lie on the sparse branches where $\rho_i$ is minimal. This wider distribution of exposure among control nodes results in only a mild AIE bias, as the contributions from highly exposed and minimally exposed control nodes partially offset. The TTE estimator remains stable, since it does not depend on neighborhood exposure and is therefore less sensitive to the clustered structure of the network.

\begin{table*}[!ht]
	\caption{Estimation results for the Ego-Facebook network}
	\label{tab: Estimation Results for Ego}	
	\centering
	\begin{tabular}{l*{6}{c}r} \hline
		Estimand & Bias \\ 
		\hline
		TTE & -0.209 \\		
		ATE & 7.095 \\
		AIE & 0.684 \\
		\hline
	\end{tabular}
\end{table*}

We further evaluate the Ego-Facebook network under a three-treatment experimental design, considering three active treatments, $t_{(0)}$, $t_{(1)}$, and $t_{(2)}$, which may correspond, for example, to advertisements, messages, and product recommendations, respectively. The optimal treatment allocation $\boldsymbol{t}^\ast$ for the three-treatment setting is presented in Figure~S4 in the Supplementary Material.
We find that nodes in the high-degree hubs and along the branches tend to share a mixture of two treatments. In contrast, low-degree nodes in sparse branches receive the same treatment.


\section{Discussion}
\label{Sec:discussion}

Experimental design for network data has received increasing attention, yet most existing approaches have been developed and evaluated using small-scale or highly simplified networks. In this paper, we address the fundamental challenge of designing experiments under interference, where classical independence assumptions such as SUTVA are violated by the underlying network structure. We propose a network-aware design framework that integrates treatment balance with network topology through an optimality criterion derived from the Fisher information matrix. By combining this criterion with a local search algorithm, the proposed approach efficiently explores the combinatorial design space and yields optimized treatment allocations for networks of practical size.

Simulation studies across a range of random graph models, including Erd\H{o}s--R\'enyi, geometric random graphs, preferential attachment, and stochastic block models, demonstrate that network geometry plays a central role in shaping optimal treatment assignments. Evaluation of causal estimands under interference reveals that while total treatment effect estimators remain relatively stable, estimators for direct and interference effects exhibit bias driven by heterogeneity in neighborhood exposure. These findings underscore the importance of incorporating network structure into both the design and analysis of experiments. Applications to real-world networks, including college housing and ego-Facebook graphs, confirm the practical utility of the proposed framework, with optimized allocations distributing treatments broadly while respecting local connectivity.

Several limitations of this work merit discussion. First, although the proposed optimization procedure efficiently explores a large combinatorial design space, it is heuristic in nature and produces near-optimal solutions rather than guaranteeing global optimality \citep{Cook01081980,Koutra_2021}. Second, our analysis is confined to undirected and static networks, whereas many real-world networks are directed, weighted, or dynamic, features that may substantially affect interference structures and treatment propagation. Third, when extending the framework to settings with multiple treatments, standard causal estimands such as TTE, ATE, and AIE are no longer well defined, as they are intrinsically tied to the conventional treatment-control paradigm. 

These limitations suggest several directions for future research. Developing optimization methods with tighter approximation guarantees or exact solutions for restricted network classes would help better characterize the optimality gap of the proposed approach. Extending the framework to directed, weighted, or dynamic networks would broaden its applicability and allow more realistic modeling of interference and treatment propagation. Finally, developing new causal estimands and corresponding inference procedures for multi-treatment network experiments remains an important open problem.


\clearpage
\appendix

\renewcommand{\thesection}{S\arabic{section}}
\renewcommand{\thefigure}{S\arabic{figure}}
\renewcommand{\thetable}{S\arabic{table}}
\renewcommand{\theequation}{S\arabic{equation}}

\renewcommand{\theHfigure}{SuppFig\arabic{figure}}
\renewcommand{\theHtable}{SuppTable\arabic{table}}
\renewcommand{\theHsection}{SuppSection\arabic{section}}

\setcounter{section}{0}
\setcounter{figure}{0}
\setcounter{table}{0}
\setcounter{equation}{0}

\section*{Supplementary Material}

\section{Random Networks}

\begin{figure*}[!ht]
     \centering
     \begin{tabular}{cc}
      \includegraphics[width=0.40\textwidth] {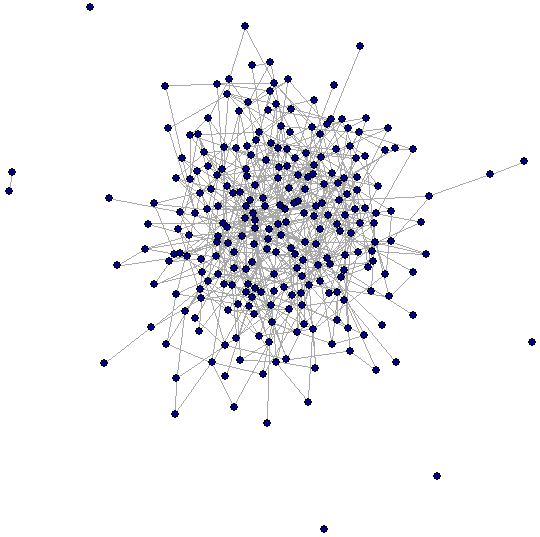} &
      \includegraphics[width=0.30\textwidth]{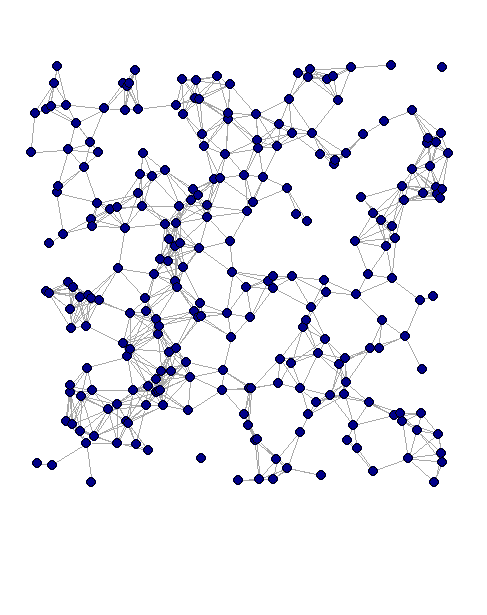} \\
       (a) ER graph. & (b) GR graph. \\
    \multicolumn{2}{c}{
     \includegraphics[width=0.32\textwidth]{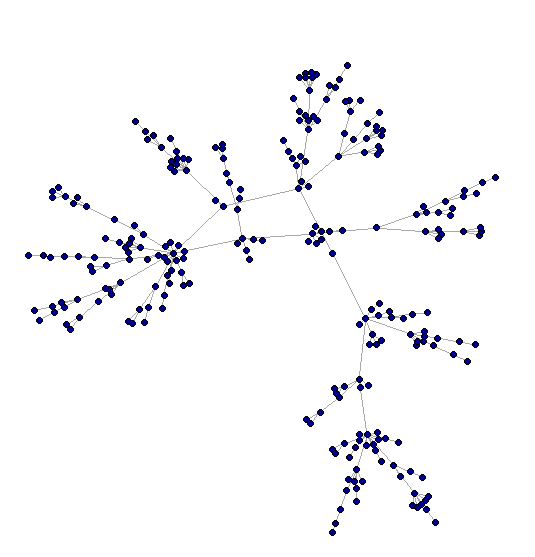}
     }\\ 
      \multicolumn{2}{c}{(c) PA model.} \\
      \end{tabular}
	\caption{Random networks each with $n=250$ nodes.}
        \label{fig:RG}
\end{figure*}

\begin{figure*}[!ht]
     \centering
     \begin{tabular}{cc}
      \includegraphics[width=0.33\textwidth]{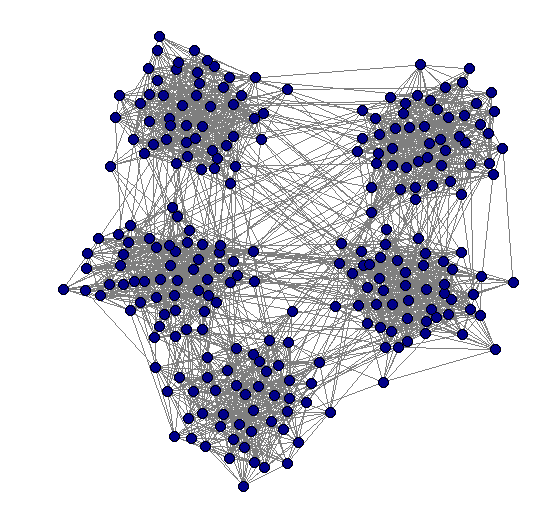} &
      \includegraphics[width=0.35\textwidth]{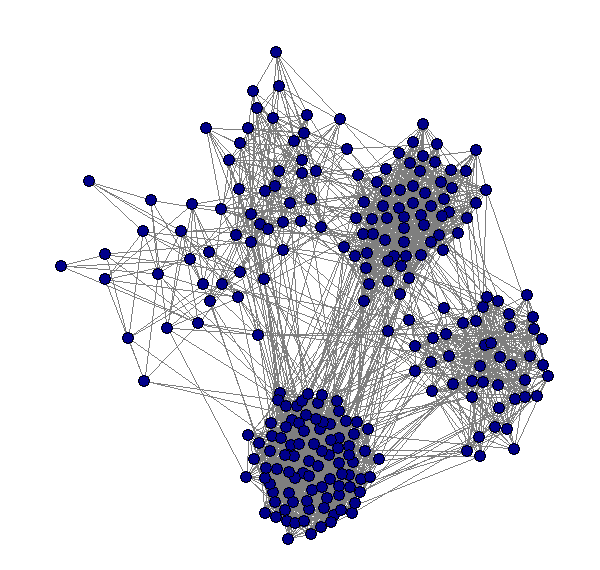} \\
       (a) Balanced. & (b) Unbalanced. \\
      \end{tabular}
	\caption{SBMs each with $n=250$ nodes.}
        \label{fig:SBM}
\end{figure*}

Figure~\ref{fig:RG} and Figure~\ref{fig:SBM} represent the random networks used in the simulation experiments. 

\newpage

\section{College housing network}

\begin{figure*}[!ht]
\centering
\includegraphics[width=0.5\linewidth]{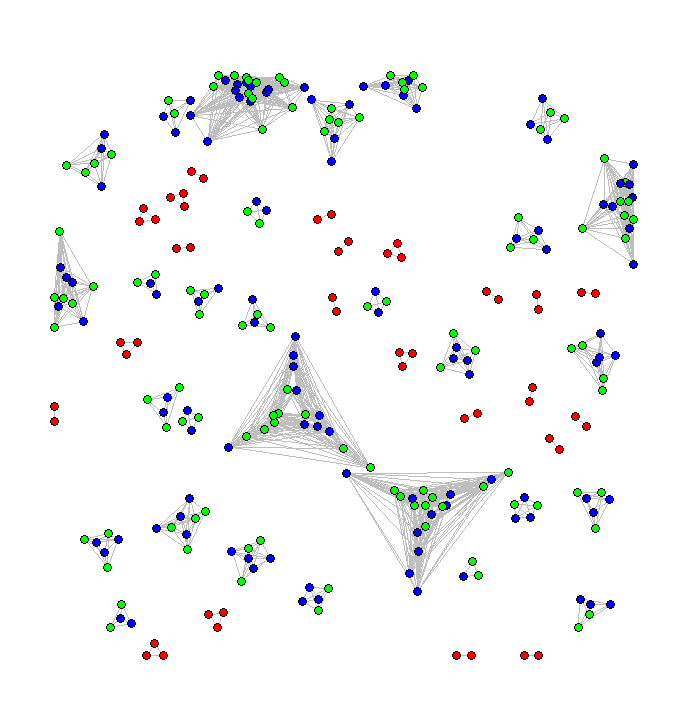}
\caption{Optimal treatment assignment for College housing network with three treatments.}
\label{fig:CN t=3}
\end{figure*}

Figure~\ref{fig:CN t=3} shows the optimal allocation of three treatments $t_{(0)}$, $t_{(1)}$, and $t_{(2)}$, where red, green and blue nodes represent $t_{(0)}$, $t_{(1)}$, and $t_{(2)}$ respectively.

\newpage

\section{Ego-Facebook network}

\begin{figure*}[!ht]
\centering
\includegraphics[width=0.5\linewidth]{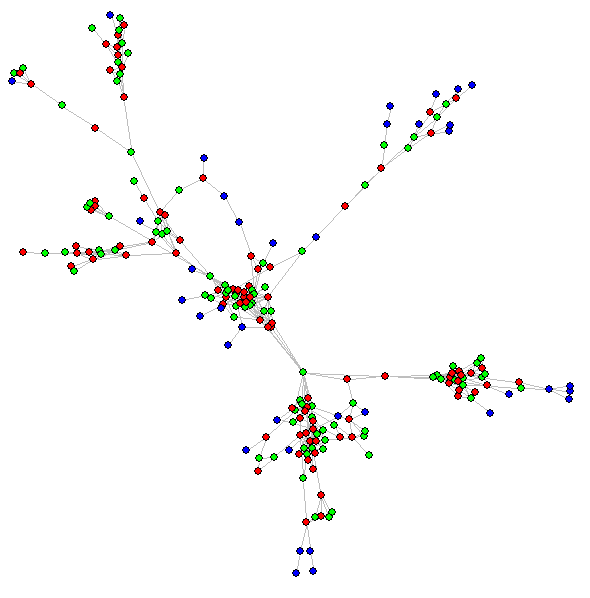}
\caption{Optimal treatment assignment for Ego-Facebook network with three treatments.}
\label{fig:Ego t=3}
\end{figure*}

Figure~\ref{fig:Ego t=3} illustrates the optimal three-treatment configuration, where nodes are assigned to $t_{(0)}$, $t_{(1)}$, and $t_{(2)}$ are colored red, green, and blue, respectively.

\bibliographystyle{unsrtnat}
\bibliography{references}  






\end{document}